\documentclass[8.5pt,twoside,twocolumn]{article}
\usepackage[super,sort&compress,comma]{natbib}
\usepackage{mhchem}
\usepackage{times,mathptm}
\usepackage{sectsty}
\usepackage{balance}
\usepackage{amssymb}

\usepackage{graphicx} 
\usepackage{lastpage}
\usepackage[format=plain,justification=raggedright,singlelinecheck=false,font=small,labelfont=bf,labelsep=space]{caption}
\usepackage{fancyhdr}
\begin{document}

\thispagestyle{plain}
\fancypagestyle{plain}{
\renewcommand{\headrulewidth}{1pt}}
\renewcommand{\thefootnote}{\fnsymbol{footnote}}
\renewcommand\footnoterule{\vspace*{1pt}%
\hrule width 3.4in height 0.4pt \vspace*{5pt}}
\setcounter{secnumdepth}{5}

\makeatletter
\def\subsubsection{\@startsection{subsubsection}{3}{10pt}{-1.25ex plus -1ex minus -.1ex}{0ex plus 0ex}{\normalsize\bf}}
\def\paragraph{\@startsection{paragraph}{4}{10pt}{-1.25ex plus -1ex minus -.1ex}{0ex plus 0ex}{\normalsize\textit}}
\renewcommand\@biblabel[1]{#1}
\renewcommand\@makefntext[1]%
{\noindent\makebox[0pt][r]{\@thefnmark\,}#1}
\makeatother
\renewcommand{\figurename}{\small{Fig.}~}
\sectionfont{\large}
\subsectionfont{\normalsize}

\fancyfoot{}
\fancyfoot[RO]{\footnotesize{\sffamily{1--\pageref{LastPage} ~\textbar  \hspace{2pt}\thepage}}}
\fancyfoot[LE]{\footnotesize{\sffamily{\thepage~\textbar\hspace{3.45cm} 1--\pageref{LastPage}}}}
\fancyhead{}
\renewcommand{\headrulewidth}{1pt}
\renewcommand{\footrulewidth}{1pt}
\setlength{\arrayrulewidth}{1pt}
\setlength{\columnsep}{6.5mm}
\setlength\bibsep{1pt}

\twocolumn[
  \begin{@twocolumnfalse}
\noindent\LARGE{\textbf{Rydberg rings}}
\vspace{0.6cm}

{\bf\noindent\large{Beatriz Olmos \textit{$^{a}$} and Igor Lesanovsky\textit{$^{a}$}}}\vspace{0.5cm}



\vspace{0.6cm}

\noindent \normalsize{Atoms in highly excited Rydberg states exhibit remarkable properties and constitute a powerful tool for studying quantum phenomena in strongly interacting many-particle systems. We investigate alkali atoms that are held in a ring lattice and excited to Rydberg states. The system constitutes an ideal model system to study thermalization of a coherently evolving quantum many-particle system in the absence of a thermal bath. Moreover, it offers exciting perspective to create entangled many-body quantum states which can serve as a resource for the generation of single photons.}
\vspace{0.5cm}
 \end{@twocolumnfalse}
  ]

\footnotetext{\textit{$^{a}$~School of Physics and Astronomy, University of
Nottingham, Nottingham, NG7 2RD, UK; E-mail: igor.lesanovsky@nottingham.ac.uk}}

\section{Introduction}
During the last decades, ultra cold atomic physics has experienced a tremendous boost initiated by the invention of laser cooling \cite{Metcalf99}. In this course, and in particular with the experimental achievement of a Bose-Einstein Condensate, it was quickly realized that ultra cold atomic systems have the potential to explore phenomena across many areas of modern physics. This is rooted in the fact that these systems are extremely versatile for their interaction properties as well as their confinement can be tailored by optical and magnetic fields to great precision \cite{Bloch08}. As a consequence, many model systems that served our understanding of many-body phenomena in condensed matter physics became suddenly realizable and could be studied experimentally. Prime examples, to name only a few, are the experimental investigation of the Mott-Insulator transition \cite{Greiner02}, the Beresinskii-Konsterlitz-Thouless phase transition in a two-dimensional bose gas \cite{Hadzibabic06}, or the recently observed pinning quantum phase transition in a Luttinger liquid \cite{Haller10}. These and further experiments have had a huge impact on the advancement of modern physics across area boundaries and have stimulated further work in condensed matter physics, quantum optics, quantum information, ultra cold chemistry and statistical mechanics.

\subsection{Rydberg atoms - Excited states with exciting perspectives}
At present the majority of ultracold atoms experiments is carried out with ground state atoms. Very recently, however, there is a growing initiative towards exploiting the unique properties of atoms in highly excited states. These so-called Rydberg atoms are blessed with remarkable properties \cite{Gallagher84}. Firstly, albeit highly excited, their lifetime can reach tens or even hundreds of microseconds. This is to be contrasted with the typical lifetime of the first excited state of alkali metal atoms which is on the order of several tens of nanoseconds.
Secondly, atoms in Rydberg states interact strongly over large distances - the interaction strength can reach several tens of MHz at a distance of the order of a micrometer \cite{Marinescu95}. Clearly, the involved timescales and interaction strengths are vastly different from those encountered in 'traditional' ultra cold atoms setups where the interatomic interaction is well-approximated by a contact potential and interaction energies are of a few kHz.
These hugely different properties illustrate that many-body physics with Rydberg atoms takes place in a completely different parameter regime, which is usually referred to as 'frozen gas' \cite{Mourachko98,Anderson98}. Here the quantum dynamics occurs in the internal atomic degrees of freedom - the electrons - while the atoms itself can be regarded as being fixed in space.

Experimental studies of this excitation dynamics have been performed by a number of experimental groups; predominantly in setups where Rydberg states were excited from an ultra cold gas of alkali metal atoms. All these experiments show a dramatic reduction of the fraction of excited atoms once the atomic density or the interaction strength surpassed a certain value \cite{Tong04,Singer04}. This is a manifestation of a central hallmark in many-particle Rydberg physics - the 'Rydberg blockade' \cite{Lukin01,Jaksch00}.
This blockade originates from the strong interaction among Rydberg atoms which inhibits the simultaneous laser excitation of two nearby atoms to Rydberg states due to the induced level shifts. It is this effect which is responsible for the rich dynamical behavior and collective character \cite{CubelLiebisch05,Heidemann07,Reetz-Lamour08,Pritchard10} of ensembles of Rydberg excited atoms. In very recent experiments the Rydberg blockade has been experimentally demonstrated for two atoms which were held in separate optical traps \cite{Urban08,Gaetan08}. Moreover, the implementation of coherent quantum operations between two such atoms has been successfully shown \cite{Isenhower10,Wilk10}. This supported, in an impressive fashion, the feasibility of various theoretical proposals that use Rydberg atoms for the implementation of quantum information processing schemes \cite{Saffman09-2}, the study of the evolution of strongly correlated quantum systems \cite{Weimer08,Olmos09-2,Lesanovsky10}, coherent quantum state preparation \cite{Olmos09-3,Pohl10,Schachenmayer10} and quantum simulation \cite{Weimer10}.

In a very recent development, Rydberg states are about to find their way also into conventional ultra cold atoms experiments as their properties can be used to create long-ranged interactions \cite{Pupillo10}. In order to overcome the above-mentioned mismatch between lifetime and interaction energy Rydberg states are merely weakly admixed by a far off-resonant laser \cite{Mayle09}. The thereby created interatomic interactions will potentially permit the observation of exotic quantum phases such as supersolids \cite{Henkel10}.

\subsection{Outline}
\begin{figure}[h]
\centering
  \includegraphics[height=6cm]{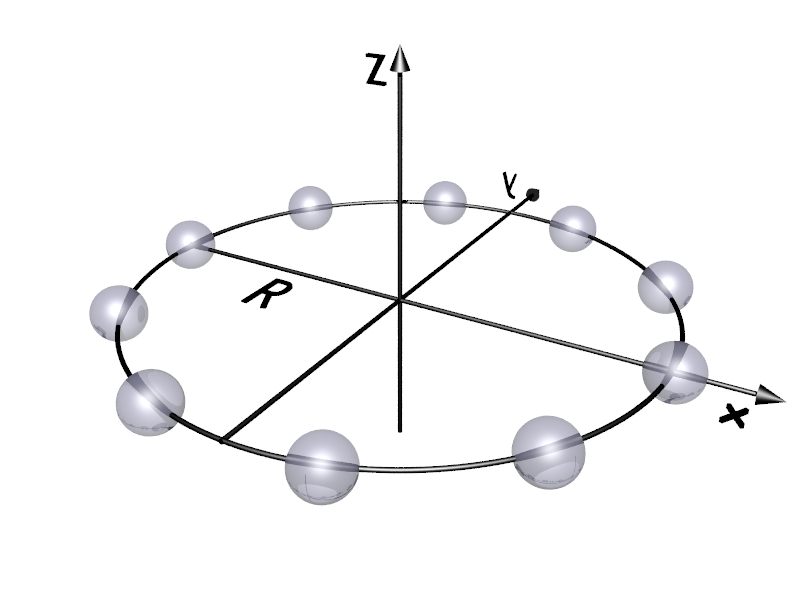}
  \caption{In our setup the atoms are held in $L$ tightly confining traps which are regularly placed on a ring with radius $R$. The typical distance between the traps is a few micrometers.}
  \label{fig:lattice}
\end{figure}

In this work we will review recent and new theoretical work carried out on so-called 'Rydberg rings'. In the system that is central to this work, ultra cold atoms are confined to tight traps which are arranged on a ring as shown in Fig. \ref{fig:lattice}.
Upon laser excitation the atoms interact strongly and the subsequent real time dynamics as well as the static properties will be thoroughly discussed in this paper. The system constitutes a paradigm example for the versatility of the Rydberg atoms. We will demonstrate that the system is ideally suited to study the transition of a closed quantum many-particle system into 'thermal equilibrium' \cite{Olmos09-2,Lesanovsky10}. Understanding the general mechanism that underlies the the thermalization of closed many-particle systems is currently a very active direction of research \cite{Kinoshita06,Hofferberth07,Rigol08,Biroli09,Pal10,Canovi10}.
Moreover, we will show that it allows to create entangled many particle states \cite{Olmos09-3,Saffman09-2,Olmos10-1}  which can serve as resource for the creation of single-photon light sources \cite{Saffman02,Porras08,Olmos10-2,Pedersen09,Nielsen10}.

\section{The Rydberg Ring}
\subsection{The ring lattice}
Our system is formed by bosonic atoms confined to a ring shaped one-dimensional lattice. The $L$ lattice sites are well-separated - with inter-site spacing $a$ - so that the low-lying quantum states of each site can be approximated by eigenstates of a local harmonic oscillator potential. The lattice spacing is of the order of several micrometers which in principle allows for single site resolution with moderate experimental effort. Such lattice can be realized by using deep large spacing optical \cite{Nelson07} or magnetic lattices \cite{Hezel06,whitlock09}.
Throughout this work we will focus on a regime in which on each site the atoms populate the harmonic oscillator ground state, whose spatial width $\sigma$ is much smaller than the lattice spacing $a$. In fact, we will assume that the atoms are infinitely localized, i.e., $\sigma/a\rightarrow 0$. Moreover, we will consider that each site contains an identical number of atoms which we denote by $N_0$.
This very idealized situation can only be approximately achieved in experiment. A fluctuating particle number and a finite width of the localized atomic wave packets caused by the quantum uncertainty, interactions and a finite temperature can in principle be accounted for. All these effects introduce disorder in the system, which slightly alters its properties \cite{Olmos09-3}. This is a very exciting topic by itself and will be expanded on elsewhere.

\subsection{Level structure and interaction}
Let us now focus on the electronic level structure of the trapped atoms. In this work we consider Rydberg states of alkali metal atoms whose level structure is particularly simple and strongly reminiscent of the hydrogen atom \cite{Gallagher84}. States with low angular momentum, however, are - unlike in the hydrogen atom - energetically well separated from a highly degenerate manifold of levels that is formed by states with high angular momentum.

In experiment Rydberg states are usually excited by a two-photon transition (see e.g. \cite{Heidemann07,Reetz-Lamour08}). Here the ground state of the atom $\left|g\right>$ is coupled via an intermediate $p$-state to a Rydberg state $\left|r\right>$. Under certain conditions which are usually satisfied experimentally the intermediate state can be eliminated and the two-photon transition is effectively replaced by a single fictitious laser \cite{Mayle09}. Due to the selection rules for dipole radiation only $s$ or $d$ Rydberg states can be excited this way.
In this work we choose $\left|r\right>$ to be an $s$-state, for these states possess two peculiar properties. First, they are energetically well isolated. Second, atoms excited to these states are, to a very good approximation, interacting via the isotropic van-der-Waals interaction. This interaction is of the form $V_\mathrm{vdW}(\mathbf{r})=C_6\times |\mathbf{r}|^{-6}$ where $\mathbf{r}$ is the interatomic separation and $C_6$ parameterizes the interaction strength. Of importance here is the scaling of $C_6$ with the degree of excitation which goes with the eleventh power of the principal quantum number \cite{Marinescu95,Singer05}. As a consequence the van-der-Waals interaction, which is ubiquitous among atoms, is strongly exaggerated when Rydberg states are involved.

\begin{figure}[h]
\centering
  \includegraphics[height=6cm]{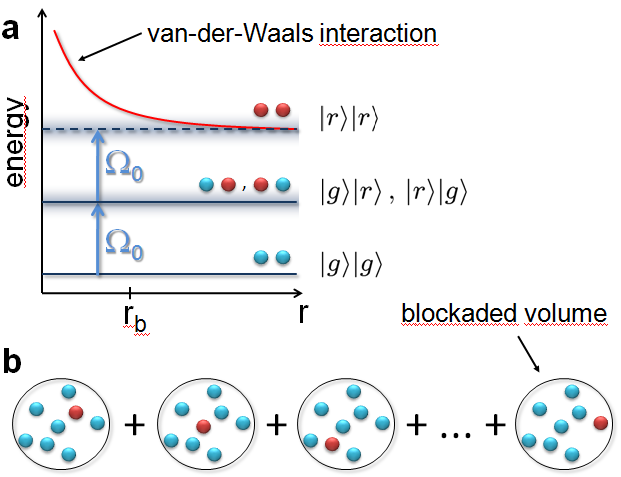}
  \caption{\textbf{a}: Level scheme for two interacting Rydberg atoms where the level structure of each atom is approximated by the two states $\left|g\right>$ and $\left|r\right>$. If the distance between the atoms is smaller than the blockade radius $r_\mathrm{b}$ the doubly excited state is no longer accessible (see text). \textbf{b}: If an ensemble of atoms is confined to a spatial region in which the distance between all atoms is smaller than $r_\mathrm{b}$ the laser can only excite a single atom at a time. This creates a superposition of all possible possible states that carry a single excitation, often referred to as \emph{superatom}.}
  \label{fig:blockade}
\end{figure}
This interaction produces a severe alteration of the excitation properties of atoms confined to a single site of the lattice which is known as the Rydberg blockade \cite{Jaksch00,Lukin01}. To understand this let us consider the situation depicted in Fig. \ref{fig:blockade}a, i.e., two atoms separated by a distance $\mathbf{r}$ whose states $\left|g\right>$ and $\left|r\right>$ are coupled resonantly via a laser with Rabi frequency $\Omega_0$.
The energy of the doubly excited states $\left|r\right>_1\left|r\right>_2$ is twice the atomic transition energy plus the interaction energy $V(\mathbf{r})=C_6\times |\mathbf{r}|^{-6}$.
This additional energy has to be overcome for the simultaneous excitation of two Rydberg atoms. This can only happen if the line corresponding to this transition (sketched by the blurred level in the figure) is sufficiently broad. For a strong laser the line width is determined by the Rabi frequency. This means in turn that atoms cannot be simultaneously excited if their separation is smaller than
\begin{eqnarray}
  r_\mathrm{b}\sim \left[C_6\,\Omega_0^{-1}\right]^\frac{1}{6},
\end{eqnarray}
i.e., if the interaction induced energy shift is larger than the Rabi frequency. The quantity $r_\mathrm{b}$ is called the \textit{blockade radius}. Throughout we will assume that this radius is much larger than the diameter of a single lattice site. Thus, out of of $N_0$ atoms located on a given site $k$ only a single one can be excited, which is sketched in Fig. \ref{fig:blockade}b. This means that on each site only two (collective) states are accessible, i.e.,
\begin{eqnarray}
\left|G\right>_k&=&\left[\left|g\right>_k\right]_1\otimes\dots\left[\left|g\right>_k\right]_{N_0}\\
\left|R\right>_k&=&\frac{1}{\sqrt{N_0}}{\cal S} \left\{\left[\left|r\right>_k\right]_1\otimes\left[\left|g\right>_k\right]_2\otimes\dots\left[\left|g\right>_k\right]_{N_0}\right\},\label{eqn:collective_states}
\end{eqnarray}
where $\cal S$ is the symmetrization operator. The state $\left|R\right>_k$ is sometimes called \textit{superatom} \cite{Heidemann07} and consists of a symmetric superposition of all possible single Rydberg excitations.

Let us now look in more detail on the excitation behavior of such a single blockaded site. The action of a laser with single atom Rabi frequency $\Omega_0$ and detuning $\Delta$ is governed by the Hamiltonian
\begin{equation*}
H_0=\sum_{k=1}^L\left[ \Omega_0\,\left|r\right>{}_k\left<g\right|+\frac{\Delta}{2}\left|r\right>{}_k\left<r\right|+\mathrm{h.c.}\right].
\end{equation*}
Having only two possible states on each site reduces the problem to that of a spin system where $\left| G\right>_k=\left| \downarrow\right>_k$ and $\left| R\right>_k=\left| \uparrow\right>_k$. Thus, in the restricted subspace spanned by the collective states (\ref{eqn:collective_states}), the Hamiltonian becomes
\begin{equation}
H_\mathrm{laser}=\sum_{k=1}^L \left[\Omega\left(\sigma^{(k)}_+ +\sigma^{(k)}_-\right) + \Delta n_k\right]
\label{eqn:laser_ham},
\end{equation}
where $\sigma^{(k)}_{\pm}=(1/2)[\sigma^{(k)}_x \pm i\sigma^{(k)}_y]$ and $n_k=\sigma^{(k)}_+\sigma^{(k)}_-$, with $\sigma_x^{(k)}$, $\sigma_y^{(k)}$ and $\sigma_z^{(k)}$ being the Pauli spin matrices. The laser coupling is strongly enhanced due to the collective nature of the state $\left| R\right>_k=\left| \uparrow\right>_k$. The corresponding \emph{collective Rabi frequency} is given through $\Omega=\Omega_0\sqrt{N_0}$.

We are now in position to formulate the spin Hamiltonian that governs the excitation dynamics of the Rydberg ring. Under the assumption that the atoms are strongly localized at the lattice sites the interaction between nearest neighbors is given by $V=C_6/a^6$. Due to the strong decay of the interaction with distance, it is sufficient to consider only this nearest neighbor interaction. This is a valid approximation provided that the interaction is not so strong that the blockade radius encompasses next-nearest neighbor sites, i.e., $\Omega \geq V/64$.
The Hamiltonian of the system then becomes
\begin{eqnarray}
  H=H_\mathrm{laser} + H_\mathrm{int}\label{eq:model_hamiltonian}
\end{eqnarray}
with the interaction Hamiltonian
\begin{equation*}
H_\mathrm{int}=V\sum_{k=1}^Ln_kn_{k+1}.
\end{equation*}
This constitutes the basis for the remainder of this paper.

Throughout we will not consider radiative decay of the Rydberg atoms. This is justified if there is a clear separation between the timescale at which the laser-driven dynamics is taking place and the atomic decay rate. In a typical experimental situation such separation is possible: The dynamics takes place on a timescale ranging from the 100 nanoseconds to a few microseconds \cite{Heidemann08} while the lifetime is on the order of 50 to 100 microseconds.

\section{Real time dynamics and thermalization}\label{sec:real_time_dynamics}
In this section we will study the real time dynamics of the model (\ref{eq:model_hamiltonian}) in the strongly interacting regime, i.e., when $V\gg \Omega$. We start from an initial state in which all atoms are in the ground state, that is, the state of the system is given by the product $\left|\mathrm{init}\right>=\prod_k \left|G\right>_k$. We will show that the long time behavior of the mean number of excitations can be understood through equilibrium thermodynamics of an equivalent classical dimer model.

\subsection{Long time behavior of observables and perfect blockade model}
\begin{figure}[h]
\centering
  \includegraphics[height=3.8cm]{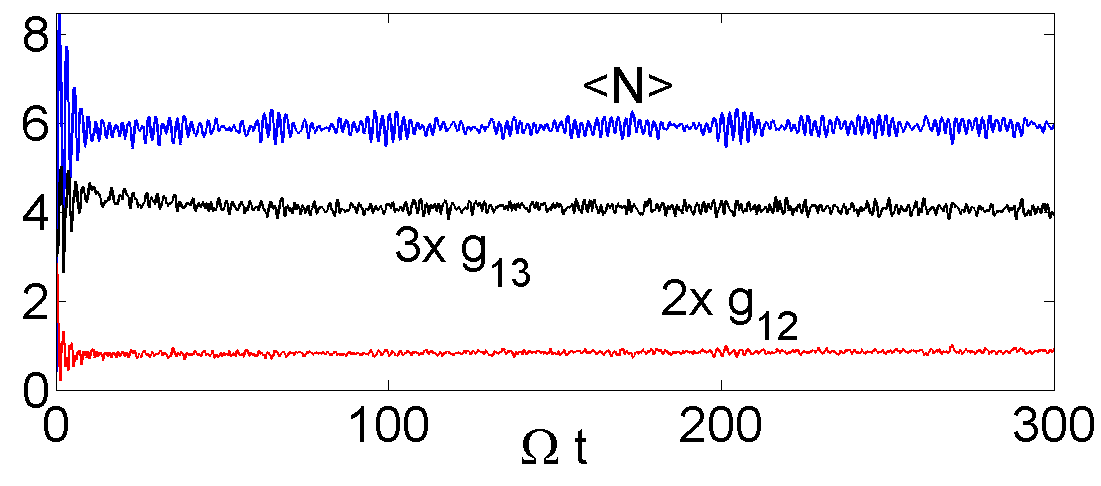}
  \caption{Temporal evolution of the mean excitation number $\left<N\right>(t)$ and the correlation functions $g_{12}(t)$ and $g_{13}(t)$ ($L=20$, $V=5\Omega$). For short times the data shows pronounced oscillations of large amplitude which are followed by a steady state regime in which small amplitude fluctuations about a stationary value take place.}
  \label{fig:eq_quantities}
\end{figure}

In Fig. \ref{fig:eq_quantities} we show the temporal evolution of the mean number of excited particles, $N=\sum_k\,n_k$ and the density-density correlations, $g_{nm}=L^2\left<n_m n_n\right>/\left<N\right>^2$ between neighboring sites and next-nearest neighbors. The data shown is in fact calculated by considering fully the $1/r^6$-tail of the van-der-Waals potential and an interaction strength $V=5\,\Omega$. In this regime the next-nearest neighbor interaction employed in eq. (\ref{eq:model_hamiltonian}) provides an excellent approximation. All three displayed quantities show similar behavior: For short times we observe a number of large contrast oscillations whose amplitude diminish for longer times, so that the quantities approach a \emph{steady state}, i.e., all of them assume a quasi time-independent value with only very small amplitude fluctuations.
Our aim is to understand the physical origin of this steady state which, however, occurs only for sufficiently strong interaction \cite{Olmos09-2,Lesanovsky10}.
To this end we employ a model which is even simpler than eq. (\ref{eq:model_hamiltonian}), called the \emph{perfect blockade} model \cite{Sun08}. Its essence is that the nearest-neighbor interaction is assumed to be infinite ($V\rightarrow \infty$). As a consequence, only the quantum states $\left|\Phi\right>$ which satisfy the constraint
\begin{eqnarray*}
  H_\mathrm{int}\left|\Phi\right>=0,
\end{eqnarray*}
are accessible by a time evolution. This in turn means that two excited atoms cannot simultaneously occupy neighboring sites, i.e., within this model the correlation function $g_{12}$ is strictly zero. For resonant laser excitation ($\Delta=0$), which we consider throughout this section, this model has no adjustable parameters and by scaling time as $t\rightarrow \Omega t$ the energy scale becomes unity. In this regard it constitutes a fundamental model for the understanding of the dynamics of a strongly interacting one-dimensional Rydberg gas.

\subsection{Evolution in excitation number space}
The strongly interacting case is characterized by a pronounced collective behavior of the system. That means the 'good' degrees of freedom or eigenexcitations of the Hamiltonian are very different from single atom excitations. Hence an understanding of the long-time behavior is most likely to be obtained by looking on the system from an angle that puts less emphasis on single atom properties.

\begin{figure}[h]
\centering
  \includegraphics[height=7cm]{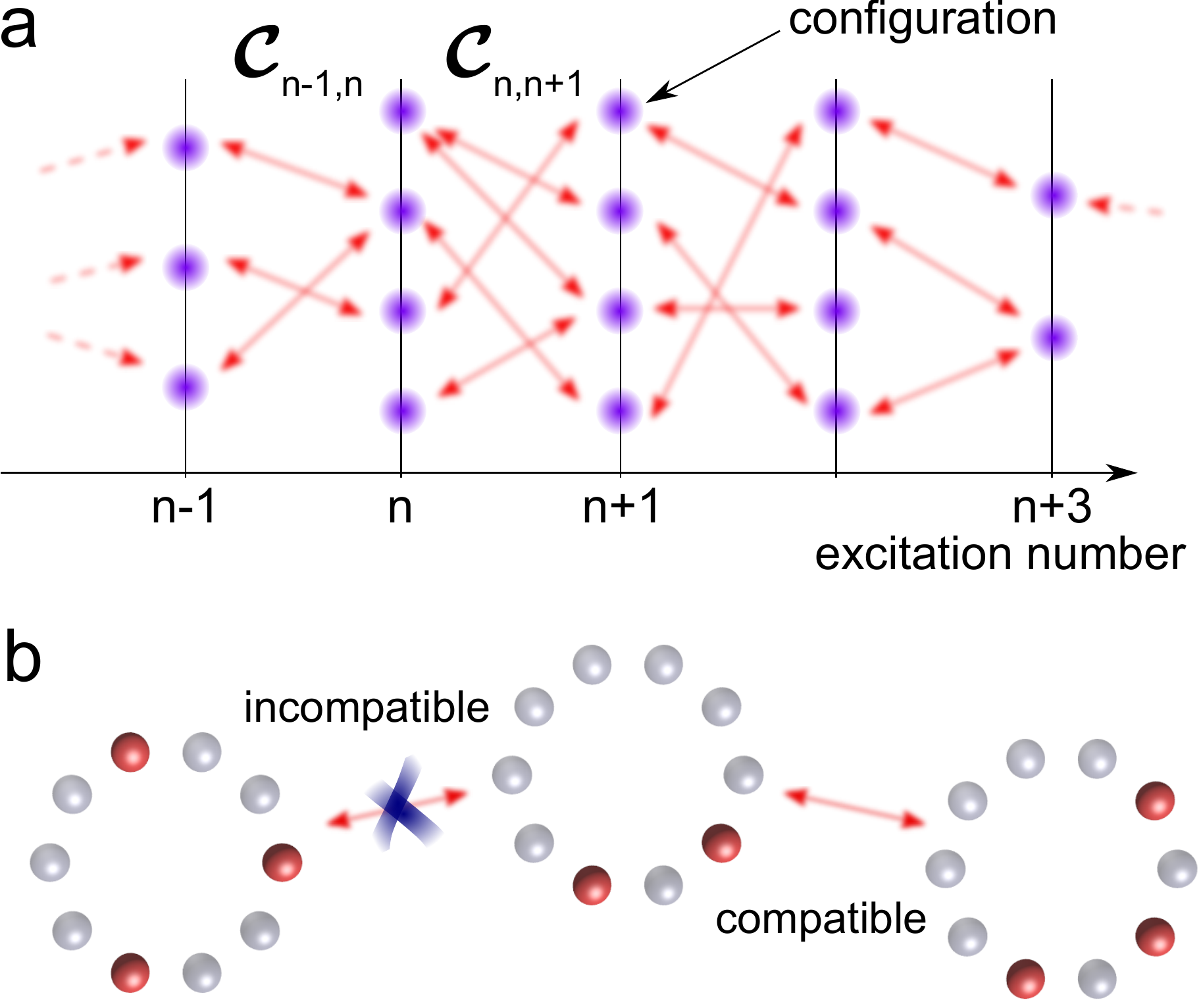}
  \caption{\textbf{a}: Sketch of the system's structure in excitation number space. The laser connects configurations whose excitation number differs by one. The precise strength of the connection is contained in the entries of the matrices $\mathcal{C}_{n,n+1}$. \textbf{b}: Transitions can only take place between compatible configurations, i.e., the laser can only connect states that can be converted into each other by the addition or removal of exactly one excitation.}
  \label{fig:excitation_number_space}
\end{figure}
Our approach to the problem of understanding of the steady state value of the mean excitation number is to study the evolution of the system in excitation number space. The underlying insight is that the configurations of the system can be characterized by two properties: Their number of excitations and the actual way these excitations are distributed on the ring. The laser can only couple configurations whose excitation number differ by one and which are compatible in the sense that the removal/addition of exactly a single excitation converts one configuration into the other. This can be cast into a diagram which is shown in Fig. \ref{fig:excitation_number_space}a. An example for a two compatible and two incompatible states is given in Fig. \ref{fig:excitation_number_space}b. Our aim is to understand the dynamics along the horizontal of this diagram as this will show us directly how the excitation number is distributed.

In the above representation the Hamiltonian acquires a block structure where each block corresponds to a different excitation number $n$. Since the laser is on resonance, all entries of each of these square blocks of dimension $\mathrm{dim}_n\times \mathrm{dim}_n$ are zero. Here, $\mathrm{dim}_n$ is the number of configurations that contain exactly $n$ excitations. The blocks are coupled through rectangular submatrices $\mathcal{C}_{n,n+1}$ with dimension $\mathrm{dim}_n\times \mathrm{dim}_{n+1}$. Thus, defining the projector on the subspace of $n$ excitations as $\left|n\right>\!\left<n\right|$ the Hamiltonian can be written as
\begin{eqnarray}
   H&=&\sum_{n,m=0}^{n_\mathrm{max}} \left|m\right>\!\left<m\right|H\left|n\right>\!\left<n\right|\nonumber\\
   &=&\sum_{n=0}^{n_\mathrm{max}}\left[ \mathcal{C}_{n,n+1}\left|n\right>\!\left<n+1\right|+\mathcal{C}^\dagger_{n,n+1}\left|n+1\right>\!\left<n\right|\right],
\end{eqnarray}
where $\left<n\right|H\left|n+1\right>\equiv\mathcal{C}_{n,n+1}$. Here $n_\mathrm{max}$ is the maximum number of excitations which is given by $L/2$ and $(L-1)/2$ for even and odd lattice size $L$, respectively.

\subsection{The rate equation}
We will now show that it is possible to derive an equation for the system's evolution in the excitation number space that possesses a steady state solution which explains the saturation of the number of Rydberg atoms observed in Fig. \ref{fig:eq_quantities}.

\begin{figure}[h]
\centering
  \includegraphics[width=9cm]{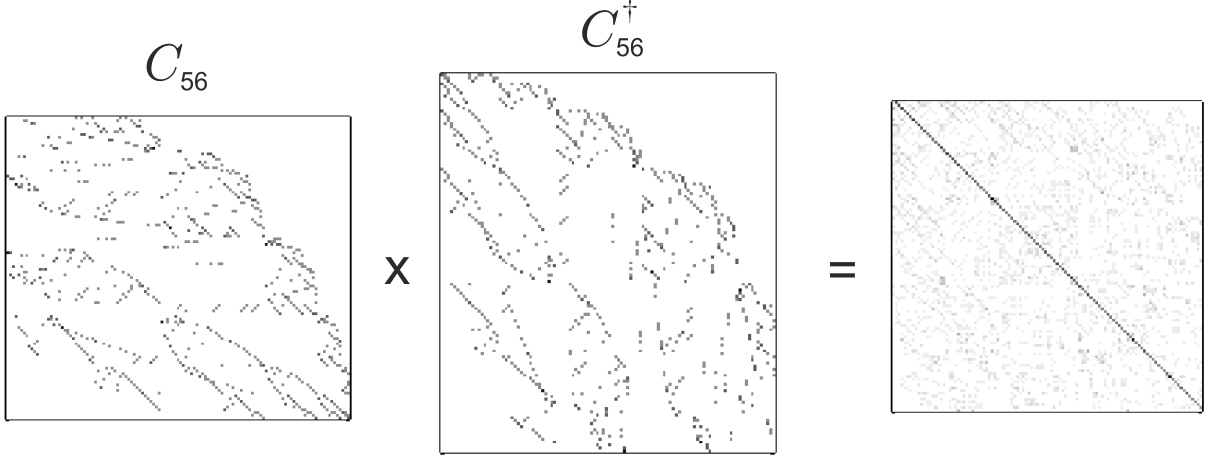}
    \caption{The product $\mathcal{C}^{}_{5,6}\mathcal{C}^\dagger_{5,6}$ ($L=20$) is approximately proportional to the identity matrix with dimension $\mathrm{dim}_5$.}
  \label{fig:c_matrices}
\end{figure}

The mean number of Rydberg atoms that are excited after a time $t$ is given by
\begin{eqnarray}
 \left<N\right>(t)=\sum_{n=0}^{n_\mathrm{max}} n\, p_n(t)\label{eq:ryd_num},
\end{eqnarray}
where $p_n(t)$ is the probability to find exactly $n$ excitations. This probability is given by $p_n(t)=\mathrm{Tr}\left[\left|n\right>\!\left<n\right|\,\rho(t)\right]$ with $\rho(t)$ being the density matrix of the system. The evolution of the density matrix is governed by the von-Neumann equation $\partial_t\rho(t)=-i\left[H,\rho(t)\right]$ which is equivalent to the integral equation $\rho(t)=\rho(0)-i \int_0^t d\tau \left[H,\rho(\tau)\right]$. Inserting this integrated equation into the von-Neumann equation and abbreviating $p_n(t)\equiv p_n$ one finds
\begin{align}
  \partial_t p_n=-i \mathrm{Tr}\left[\left[H,\rho(0)\right]\,\left|n\right>\!\left<n\right|\right]-\int_0^t d\tau\, \mathrm{Tr}\left[\left[H,\left[H,\rho(\tau)\right]\right]\,\left|n\right>\!\left<n\right|\right].
\end{align}
The first term vanishes if $\rho(0)$ is diagonal in the excitation number space, i.e., if the initial state contains a defined number of Rydberg atoms, which we assume here. The second term can be rearranged such that
\begin{align}
  \partial_t p_n=\int_0^t\!\! d\tau\, \mathrm{Tr}\left[\left(2H\left|n\right>\!\left<n\right|H-\left\{\left|n\right> \!\left<n\right|,H^2\right\}\right)\rho(\tau)\right].\label{eq:rate_equation_1}
\end{align}
Let us study the first term of this equation in order to see how it can be further simplified: Upon insertion of two complete basis sets and abbreviating $\rho_{mk}(\tau)=\left<m\right|\rho(\tau)\left|k\right>$\footnote{Note that $\rho_{mk}(\tau)$ is not a matrix element but in general (due to the large degeneracy of the excitation number subspaces) a $\mathrm{dim}_m \times \mathrm{dim}_k$-matrix.} we find
\begin{eqnarray}
  \sum_{km}\mathrm{Tr}\left[\left|k\right>\!\left<k\right|H\left|n\right>\!\left<n\right|H\left|m\right>\!\left<m\right|\rho(\tau)\right]=\sum_{km} \mathrm{Tr}\, \mathcal{C}^{\phantom{\dagger}}_{k,n}\mathcal{C}^\dagger_{m,n} \rho_{mk}(\tau).\nonumber
\end{eqnarray}
The crucial observation is now that $\mathcal{C}^{\phantom{\dagger}}_{k,n}\mathcal{C}^\dagger_{m,n}\approx \kappa_{mn} \mathbf{1}_m \delta_{km}$ where $\kappa_{mn}$ is a constant and $\mathbf{1}_m$ is the $\mathrm{dim}_m$-dimensional representation of the identity matrix that at the same time is the projector onto the subspace containing $m$ excitations, so that $\mathrm{Tr}\left[\mathbf{1}_m\rho_{mm}\right]=p_m$. The reason for this is that the matrices $\mathcal{C}^{\phantom{\dagger}}_{k,n}$ are sparsely occupied and that the entries, due to the strong interaction, are completely uncorrelated. Thus, when calculating the products $\mathcal{C}^{\phantom{\dagger}}_{k,n}\mathcal{C}^\dagger_{m,n}$, the only elements that acquire an appreciable size stem from a row of $\mathcal{C}^{\phantom{\dagger}}_{k,n}$ being multiplied by itself, i.e., the corresponding column of $\mathcal{C}^\dagger_{m,n}$. This happens only if $m=k$ and only for the diagonal elements.
This is illustrated in Fig. \ref{fig:c_matrices} where we look at the product $\mathcal{C}^{\phantom{\dagger}}_{5,6}\mathcal{C}^\dagger_{5,6}$ for $L=20$.
We thus find
\begin{eqnarray}
   \sum_{km}\mathrm{Tr}\, \mathcal{C}^{\phantom{\dagger}}_{k,n}\mathcal{C}^\dagger_{m,n} \rho_{mk}(\tau)\approx \kappa_{n-1,n}\,p_{n-1}(\tau)+\kappa_{n+1,n}\,p_{n+1}(\tau).
\end{eqnarray}
There are three more steps which we have to perform in order to arrive at a closed rate equation for the probabilities $p_n(t)$: Firstly, we treat the two remaining terms of eq. (\ref{eq:rate_equation_1}) in the same way we have treated the first term. Secondly, we make use of the fact that $\mathrm{Tr}\left[\mathcal{C}^{\phantom{\dagger}}_{m,n}\mathcal{C}^\dagger_{m,n}\right]=\mathrm{Tr}\left[\mathcal{C}^\dagger_{m,n}\mathcal{C}^{\phantom{\dagger}}_{m,n}\right]$ which fixes the constants $\kappa_{n,n+1}=u_n \mathrm{dim}_{n+1}$ and $\kappa_{n+1,n}=u_n \mathrm{dim}_{n}$ where the $u_n$ are unknown coefficients which we assume are non-zero. Thirdly, we ignore the behavior of eq. (\ref{eq:rate_equation_1}) on very short time intervals, i.e., the evolution of the system on a time-scale shorter than $\Omega^{-1}$, which is the typical timescale given by the laser. Assuming that during such a coarse grained time-interval of length $\tau$ the density matrix does not change we arrive at the following equation for the probabilities
\begin{align}
  \frac{p_n(\tau)-p_n(0)}{\tau}&=&\tau\left[u_{n-1}\mathrm{dim}_{n}p_{n-1}+u_n \mathrm{dim}_np_{n+1}\right]\nonumber\\
  &&-\tau\left[u_n \mathrm{dim}_{n+1} +u_{n-1} \mathrm{dim}_{n-1}\right]p_n.\label{eq:rate_equation}
\end{align}
A more quantitative and thorough discussion of the derivation of this equation is given in Ref. \cite{Olmos09-2}.

The steady state of this equation satisfies $\frac{p^\mathrm{steady}_n(\tau)-p^\mathrm{steady}_n(0)}{\tau}=0$ and is given by
\begin{eqnarray}
  p^\mathrm{steady}_n=\frac{\mathrm{dim}_n}{\mathrm{dim}}\label{eq:steady_state}
\end{eqnarray}
with the dimension of the total Hilbert space $\mathrm{dim}=\sum_{n=0}^{n_\mathrm{max}}\mathrm{dim}_n$. $p^\mathrm{steady}_n$ represents the probability distribution of the number of excited atoms or the full statistics of the Rydberg number count in the steady state. Remarkably it can be obtained without a precise knowledge of the numbers $u_n$.

For the Rydberg ring the steady state solution (\ref{eq:steady_state}) can be given explicitly:
\begin{eqnarray}
 p_n^\mathrm{steady}=\frac{1}{\mathrm{dim}}\,\frac{L}{L-n}\left(
 \begin{array}{c}
 L-n \\ n \\
 \end{array}
 \right)\label{eq:rho_analytical}
\end{eqnarray}
with
\begin{eqnarray}
   \mathrm{dim}=\sum_{n=0}^{n_\mathrm{max}}\frac{L}{L-n}\left(
   \begin{array}{c}
   L-n \\ n \\
   \end{array}
   \right).\label{eq:partition_function}
\end{eqnarray}
This allows us to extract the mean value of the number of Rydberg atoms in the steady state, which for $L\rightarrow\infty$ evaluates to
\begin{eqnarray}
\bar{\left<N\right>}=\frac{L}{2}\left[1-\frac{1}{\sqrt{5}}\right]\approx 0.276\,L.\label{eq:mean_rydberg_number}
\end{eqnarray}

\subsection{Numerical results}
Let us now compare our findings to the numerical solution of the system when the initial state is the product state $\left|\mathrm{init}\right>=\prod_k \left|G\right>_k$, i.e., $p_0(0)=1$.
\begin{figure}[h]
\centering
  \includegraphics[width=9cm]{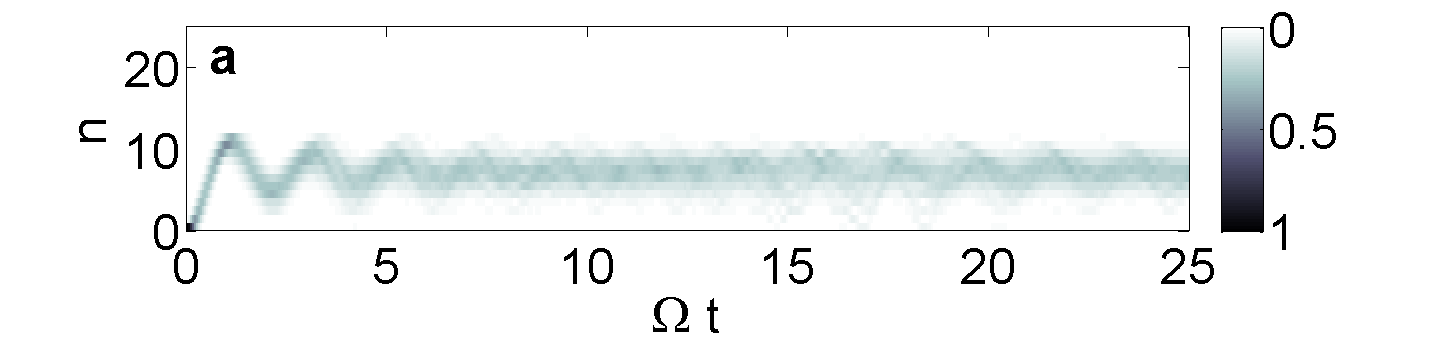}
  \includegraphics[width=9cm]{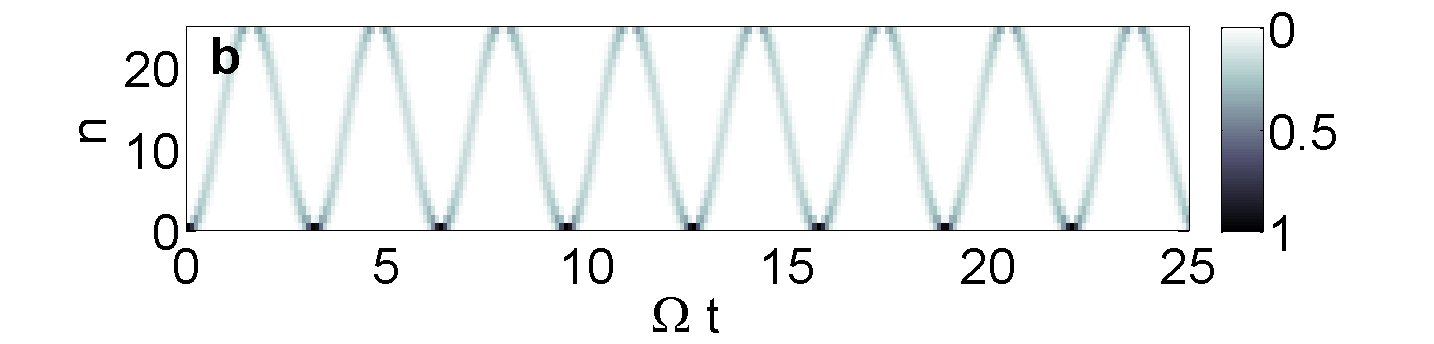}
  \caption{Temporal evolution of the probability density $p_n$ in a system consisting of $L=25$ sites in excitation number space. Initially all atoms are in the ground state, i.e. $p_0=1$. \textbf{a}: In the interacting case the system reaches eventually a state in which the probability density localizes in excitation number space. \textbf{b}: This is not the case in the absence of interactions. Here, the wave packet performs coherent oscillations with maximal amplitude.}
  \label{fig:numerics}
\end{figure}
In Fig. \ref{fig:numerics}a we show the temporal evolution of the probability density $p_n$ in excitation number space, i.e., along the horizontal axis of the diagram shown in Fig. \ref{fig:excitation_number_space}a. For short times we observe oscillations which quickly diminish after $\Omega t\approx 5$. Beyond that time the distribution becomes strongly localized and maintaining its shape with slight oscillations on top. Here the system is localized in excitation number space populating mainly configurations with excitation numbers close to $\bar{\left<N\right>}$. This relaxation occurs due to the quasi-random or uncorrelated couplings between the different excitation number subspaces which constituted the basis for the derivation of eq. (\ref{eq:rate_equation}). This eventually washes out all phase coherence between the subspaces containing a different number of excitations. In the non-interacting case this phase coherence is maintained throughout so that the population shows (Rabi-)oscillations of full contrast, as shown in Fig. \ref{fig:numerics}b.

\begin{figure}[h]
\centering
  \includegraphics[width=4cm]{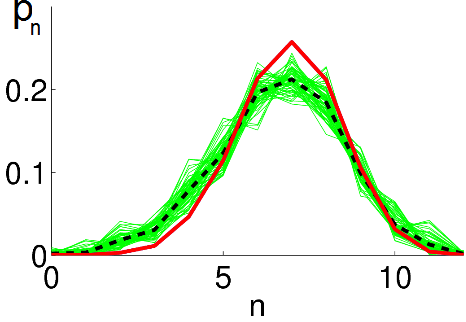}
  \caption{Probability density $p_n$ in excitation number space for $L=25$. The green (thin) curves are snapshots taken during the interval $100\leq t \leq 104$. For these times the calculated Rydberg number shows the steady state shown in Figs. \ref{fig:eq_quantities} and \ref{fig:numerics}a. The dashed curve is obtained by taking the average over the set of snapshots. The red thick curve shows $p_n^\mathrm{steady}$ as given by eq. (\ref{eq:rho_analytical}).}
  \label{fig:distribution}
\end{figure}
To compare the numerically obtained distribution to the analytically predicted one (\ref{eq:rho_analytical}) we remove the temporal fluctuations (which are a finite-size effect) by averaging the distribution over a time interval of a certain length. We here choose the interval $100\leq \Omega t \leq 104$ during which the steady state is well-established. The result is presented in Fig. \ref{fig:distribution}. The agreement is good, both distributions are peaked at the same value, i.e., $n=7$ for $L=25$. However, the numerical result has a systematically enhanced probability at low $n$. This deviation is also reflected in the mean number of Rydberg atoms which yields $\bar{\left<N\right>}_\mathrm{numerical}\approx 0.26 L$ which is slightly smaller than the result (\ref{eq:mean_rydberg_number}). The reason lies in the particular choice of the initial state. When choosing a random initial state (which might be difficult experimentally) one obtains in general a perfect agreement between the numerical data and the analytical result as is shown in Ref. \cite{Olmos09-2}.

\subsection{Connection to the classical hard core dimer model}
Having found strong evidence for the fact that our coherently driven and closed quantum system thermalizes, it is tempting to ask the question whether there exists an analogous classical model whose thermal equilibrium properties coincide with the properties of the quantum model for long times.
\begin{figure}[h]
\centering
  \includegraphics[width=8cm]{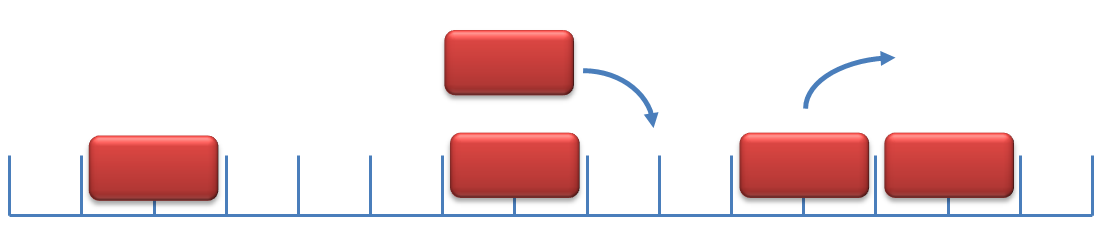}
  \caption{Classical lattice gas of hard-core dimers whose thermal equilibrium properties coincide with the steady state that is assumed by the Rydberg ring for long times. The laser adds or removes dimers which cannot simultaneously occupy the same lattice site.}
  \label{fig:lattice_gas}
\end{figure}
In our system the laser adds/removes Rydberg atoms to/from the lattice. The size of the Rydberg atoms is effectively two lattice sites and they are also \emph{hard objects} as there is strict nearest neighbor exclusion. This is reminiscent of a classical hard core dimer model as shown in Fig. \ref{fig:lattice_gas}. Let us now calculate the mean number of dimers occupying the lattice in this classical model. To this end we first assume that the dimers are not \emph{hard objects} but that the simultaneous occupation of a site by two of them costs an energy $V$. The grand canonical partition function of this system is then given by  $\Xi(\beta,\mu,L)=\left(\lambda_+\right)^L+\left(\lambda_-\right)^L$
with $\lambda_\pm \equiv \frac{1}{2} \left( 1 +e^{\beta(\mu-V)} \right) \pm \frac{1}{2} \sqrt{4 e^{\beta\mu}+\left(1-e^{\beta(\mu-V)}\right)^2}$ where $\beta$ is the inverse temperature and $\mu$ the chemical potential. The mean number of dimers in equilibrium is then given by
\begin{eqnarray}
\bar{\left<N\right>}_\mathrm{th}=\frac{L}{2}\left[1 -\frac{1-e^{-\beta V}}{ \sqrt{4+\left(e^{-\beta V}-1\right)^2}}\right]\stackrel{\beta V\rightarrow\infty}=\frac{L}{2}\left[1-\frac{1}{\sqrt{5}}\right]\label{eq:thermal_exp_value}
\end{eqnarray}
which, in the limit $\beta V\rightarrow\infty$, i.e., infinitely strong interaction, becomes the result that we have previously obtained for the number of Rydberg atoms [eq. (\ref{eq:mean_rydberg_number})]. This is not surprising because eq. (\ref{eq:partition_function}), i.e., the dimension of the Hilbert space of the perfect blockade model, is by construction the microcanonical partition function of the hard core dimer model or one–dimensional monomer-dimer problem \cite{Orlando07}.

This shows that in equilibrium the probability for the system to reside in a subspace with $n$ Rydberg atoms is determined solely by the entropy. This means that the probability $p_n$ is simply proportional to the number of available configurations or microstates that contain $n$ such excitations.

We have to emphasize at this point that the quantum state of the system at any time possesses zero entropy, even in the steady state. This is a consequence of the fact that a pure quantum state always remains a pure state under a coherent evolution. The entropy we referred to previously is the one related to the classical subsystem. This classical subsystem is characterized by a density matrix in which all the information which is not necessary for evaluating properties of classical observables - such as the mean particle number or density-density correlations - is traced out. In the steady state this density matrix has maximum entropy, i.e., it is completely mixed.

\subsection{Beyond the perfect blockade model}
We have seen that the microcanonical ensemble very accurately predicts the long time properties of the distribution of Rydberg excitations. All configurations occurred with equal weight which is a result of the fact that we chose to consider only configurations with zero interaction energy. When accounting for a finite but large nearest-neighbor interaction strength $V$, a significantly larger number of configurations becomes accessible during the dynamics. The population of configurations with high interaction energy will, however, be suppressed during the time-evolution since those are far detuned from the initial state with zero interaction. Interestingly, it turns out that in the long-time limit the properties of the system can again be understood by using statistical mechanics if the interaction strength surpasses a certain value which is $V_\mathrm{trans}\approx 2 \Omega$ \cite{Lesanovsky10}. Here, for example, the mean particle number is given by eq. (\ref{eq:thermal_exp_value}) with a finite $\beta V$. The inverse temperature $\beta$ can be determined by fitting the distribution of the interaction energies that follow a Boltzmann law, i.e., the probability $p_\epsilon$ to find the system in a state with interaction energy $\epsilon$ is $p_\epsilon\propto e^{-\beta\,\epsilon}$. Further numerical analysis shows that the quantum state at long times not only shows this characteristic distribution of the interaction energies, but also that other elementary relations known from thermal equilibrium states such as the fluctuation-dissipation theorem are satisfied \cite{Lesanovsky10}.

It will be interesting to explore this further, not only theoretically, but in particular experimentally. This will shed light on the question: \emph{Under which conditions can aspects of a closed many-body quantum system be understood by equilibrium thermodynamics?} Experiments with Rydberg atoms in large lattices will be especially useful to probe and characterize the nature of the transition occurring in the vicinity of $V_\mathrm{trans}$.

\section{Quantum state preparation}
The subject matter of this section is again the model (\ref{eq:model_hamiltonian}) but with focus on the weakly interaction regime, i.e., $V\ll \Omega$. We will see that in this regime the dynamics of the system takes place in constrained subspaces in which the quantum evolution is approximately governed by an exactly solvable Hamiltonian. The properties of the eigenenergies and the corresponding many-particle eigenstates will be analyzed and a scheme for their experimental realization will be outlined. Finding simple ways for creating entangled many-particle states is of importance for numerous applications: They serve as resource for precision quantum measurements \cite{DAriano01}, for measurement based quantum computing \cite{Briegel09}, as well as for the creation of single-photon light sources \cite{Porras08}. We will conclude by discussing in particular the realization of a single-photon source based on a delocalized excitation within the Rydberg ring.

\subsection{Constrained dynamics}
In the weakly interacting regime the laser Hamiltonian (\ref{eqn:laser_ham}) (in particular the term proportional to the Rabi frequency, since we assume $\Omega\gg|\Delta|$) dominates the dynamics. It is therefore convenient to choose a basis in which it is diagonal. This is achieved by the unitary transformation
\begin{eqnarray}
U=\prod_{k=1}^L \exp{\left(-i\frac{\pi}{4}\sigma_y^{(k)}\right)}\label{eq:rotation}
\end{eqnarray}
which brings $\sigma_x\rightarrow\sigma_z$ and $\sigma_z\rightarrow-\sigma_x$. When applied to our Hamiltonian (\ref{eq:model_hamiltonian}) we obtain
\begin{equation}\label{eqn:UHU}
H'=U^\dagger HU=\frac{V L}{4}+H_\mathrm{xy}+H_1+H_2,
\end{equation}
with
\begin{eqnarray}
H_\mathrm{xy}&=&\sum_{k=1}^L \left[\Omega\sigma_z^{(k)}+\frac{V}{4}\left(\sigma_{+}^{(k)}\sigma_{-}^{(k+1)}+\sigma_{-}^{(k)}\sigma_{+}^{(k+1)}\right)\right]\label{eqn:H_xy}\\
H_1&=&\frac{\Delta}{2}\sum_{k=1}^L\left(1-\sigma_{x}^{(k)}\right)\label{eqn:H_1}\\
H_2&=&\frac{V}{4}\sum_{k=1}^L \left[\left(\sigma_{+}^{(k)}\sigma_{+}^{(k+1)}+\sigma_{-}^{(k)}\sigma_{-}^{(k+1)}\right)-2\sigma_x^{(k)}\right]\label{eqn:H_2},
\end{eqnarray}
where $H_\mathrm{xy}$ is the well-known $xy$-model of a chain of spin $1/2$ particles in a transverse magnetic field.

\begin{figure}
\includegraphics[width=4cm]{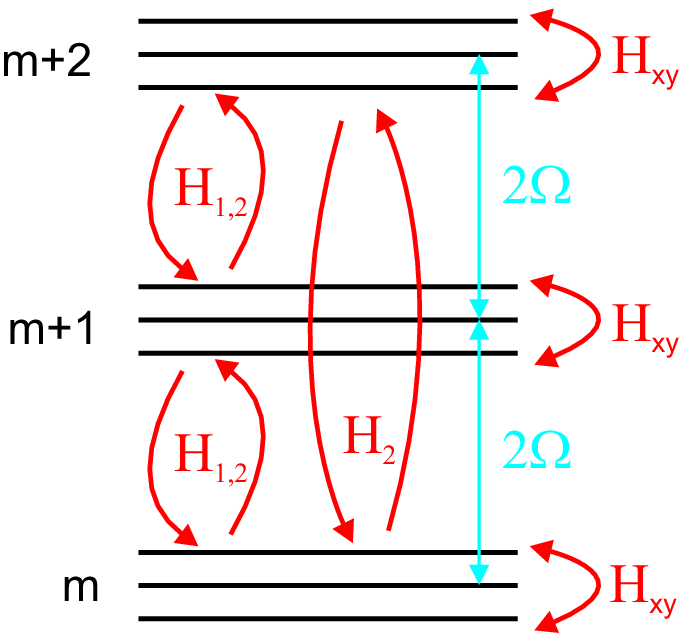}
\centering
\caption{Level structure in the regime $\Omega\gg V$ and $|\Delta| \ll \Omega$. The spectrum splits into manifolds which can be labeled by the quantum number $m$ of the operator $\sum_k \sigma^{(k)}_z$. For sufficiently large $\Omega$, the coupling between manifolds that is established only by $H_1$ and $H_2$ can be neglected. The (constrained) dynamics inside the $m$-subspaces is then solely determined by $H_\mathrm{xy}$.}\label{fig:spectrum}
\end{figure}

Let us now analyze the importance of the individual contributions to $H'$. As we can see in Fig. \ref{fig:spectrum}, the spectrum of the Hamiltonian decays into manifolds of states which are separated by gaps whose width is approximately $2\Omega$. This is caused by the dominant first term of $H_\mathrm{xy}$, i.e., $\Omega\sum_k\sigma_z^{(k)}$. The eigenstates of $\sigma^{(k)}_z$ are - in terms of the (super)atom states - given by
\begin{equation*}
\left|\pm\right>_k=\frac{1}{\sqrt{2}}U^\dagger\left[\left|G\right>_k\pm\left|R\right>_k\right]
\end{equation*}
with $\sigma^{(k)}_z\left|\pm\right>_k=\pm\left|\pm\right>_k$. Thus, each of the manifolds that determine the coarse structure of the spectrum is spanned by a set of product states that have the same number of (super)atoms in the state $\left|+\right>$. In Fig. \ref{fig:spectrum} we show the corresponding coarse level structure. There, we label each manifold by the eigenvalue of its states with respect to the operator $\sum_{k}^L\sigma_z^{(k)}$, $m$, which basically counts the difference between the number of sites that are in the $\left|+\right>$ and the $\left|-\right>$ state.

The second term of $H_\mathrm{xy}$ conserves the total number of $\left|+\right>$ (super)atoms. In other words, it couples only states that belong to the same $m$-manifold and that are nearly degenerate. As a consequence, the strength of these intra-manifold couplings due to $H_{\mathrm{xy}}$ is proportional to the interaction strength $V$. Conversely, $H_1$ and $H_2$ couple states that belong to manifolds with different number of (super)atoms in the state $\left|+\right>$.
In particular, $H_1$ and the last term of $H_2$ flip one of the (super)atoms from $\left|+\right>$ to $\left|-\right>$ or viceversa. Thus, the coupled states belong to different manifolds with $\Delta m=\pm1$, energetically separated by $2\Omega$. The two first terms of $H_2$ drive a similar process, flipping always two contiguous (super)atoms in the same state simultaneously, i.e., $\left|++\right>\rightarrow\left|--\right>$ or $\left|--\right>\rightarrow\left|++\right>$. As a result, these terms connect states with eigenvalue $m$ to those with $m\pm2$ and which possess a energetic separation of approximately $4\Omega$. These selection rules are indicated in Fig. \ref{fig:spectrum}.
The transition rates between $m$-manifolds corresponding to $H_1$ and $H_2$ can be estimated by second order perturbation theory to be of the order $\Delta^2/\Omega$ and $V^2/\Omega$, respectively. Hence, for sufficiently strong driving $\Omega\gg V$, their contribution can be neglected and the system's dynamics is constrained to the $m$-manifolds. As a consequence, the Hamiltonian that drives the intra-manifold dynamics, given by $H_\mathrm{xy}$, effectively drives the dynamics of the \emph{entire system} in the considered parameter regime. This Hamiltonian is analytically solvable, and we thus have access to the actual spectrum and eigenstates of the system.

\subsection{Many-particle eigenstates}
We will now solve the Hamiltonian $H_\mathrm{xy}$ in order to determine the eigenstates and eigenenergies. This is done using the so-called Jordan-Wigner transformation and a subsequent Fourier transform \cite{DePasquale09}.

The Jordan-Wigner transformation introduces the operators
\begin{equation}\label{eqn:J-W}
c_k^{\dagger}=\sigma_+^{(k)}\prod_{j=1}^{k-1}\left(-\sigma_z^{(j)}\right)\quad c_k=\prod_{j=1}^{k-1}\left(-\sigma_z^{(j)}\right)\sigma_-^{(k)},
\end{equation}
which are highly non-linear in the spin operators and obey the canonical fermionic algebra
\begin{eqnarray*}
\{c^\dagger_i,c_j\}=\delta_{i,j}\qquad\{c^\dagger_i,c^\dagger_j\}=\{c_i,c_j\}=0.
\end{eqnarray*}
After this transformation, the Hamiltonian (\ref{eqn:H_xy}) takes on the form
\begin{eqnarray}\label{eqn:H_JW}\nonumber
H_{\mathrm{xy}}&=&\sum_{k=1}^L \left[2\Omega\left(c_k^\dagger c_k-\frac{1}{2}\right)+\frac{V}{4}\left(c_k^\dagger c_{k+1}+c_{k+1}^\dagger c_k\right)\right]\\
&-&\frac{V}{4}\left(c_L^\dagger c_1+c_1^\dagger c_L\right)\left(e^{i\pi n_+}+1\right).
\end{eqnarray}
Hence, the Hamiltonian has been transformed into one which describes a chain of spinless fermions with nearest neighbor hopping. The last term of Hamiltonian (\ref{eqn:H_JW}) appears due to the periodic boundary conditions. It depends on the operator $n_+=\sum_{j=1}^{L}c_j^\dagger c_j$ which counts the total number of fermions, which is also equivalent to the number of sites in the state $\left|+\right>$. We thus have to explicitly distinguish between an odd or even number of fermions (which we refer to as parity). With this distinction being manifest, $H_\mathrm{xy}$ reads
\begin{eqnarray*}
H^\mathrm{(e/o)}_\mathrm{xy}&=&\sum_{k=1}^L 2\Omega\left(c_k^\dagger c_k-\frac{1}{2}\right)+\frac{V}{4}\sum_{k=1}^{L-1}\left(c^\dagger_{k}c_{k+1}+c^\dagger_{k+1}c_k\right)\\
&&\mp\frac{V}{4}\left(c^\dagger_{L}c_{1}+c^\dagger_{1}c_L\right),
\end{eqnarray*}
for even (e) or odd (o) parity, respectively.

We proceed by introducing the Fourier transformed operators $\eta^\dagger_{p,n}=\frac{1}{\sqrt{L}}\sum_{k=1}^L \exp\left(i\alpha^p_n k\right)c^\dagger_k$ with the index $p=\mathrm{o}/\mathrm{e}$ denoting the parity and with $\alpha^\mathrm{o}_n=2n\pi/L$ and $\alpha^\mathrm{e}_n=(2n-1)\pi/L$. This finally leads to the diagonal representation of the Hamiltonian:
\begin{eqnarray}
  H^p_\mathrm{xy}=-L\Omega+\sum_{n=1}^L\eta^\dagger_{p,n}\eta_{p,n}\left(2\Omega+\frac{V}{2}\cos\alpha^p_n\right).\label{eq:fermion_hamiltonian_omega}
\end{eqnarray}
The ground state is given by
\begin{eqnarray}
  \left|0\right>=\prod_{k=1}^L \left|-\right>_k\label{eq:ground_state}
\end{eqnarray}
and the excited states are formed by the successive application of the creation operators on the ground state. In particular, the singly and doubly excited states are of the form
\begin{eqnarray}
  \left|j\right>&=&\eta_{\mathrm{o},j}^\dagger\left|0\right>\nonumber\\
  \left|ij\right>&=&\eta_{\mathrm{e},i}^\dagger\eta_{\mathrm{e},j}^\dagger\left|0\right>,\label{eq:excited_states}
\end{eqnarray}
respectively.
Note that the notion 'singly and doubly excited' does not mean that one or two Rydberg atoms are excited. It means that one or two out of the $L$ atoms on the ring are in the state $\left|+\right>$ while all others are in $\left|-\right>$. This means that the mean number of Rydberg atoms forming all the above-mentioned many particle states is $L/2$.

\subsection{Experimental preparation}
Our aim is now to find a scheme which allows to populate the excited many-body quantum eigenstates of the system in an experiment. In the same spirit as in Sec. \ref{sec:real_time_dynamics}, we assume that the initial condition is such that no Rydberg atom is present and the system is thus prepared in the state $\left|\mathrm{init}\right>=\prod_k \left|G\right>_k$. The proposed scheme consists of two steps, the preparation of the ground state (\ref{eq:ground_state}) of the Hamiltonian (\ref{eq:fermion_hamiltonian_omega}) and the subsequent excitation of the states (\ref{eq:excited_states}).

\subsubsection{Preparation of the ground state of $H_\mathrm{xy}$: }
In the following we will consider the simple situation in which there is only a single atom per lattice site. In this case the superatom state $\left|R\right>_k$ becomes simply the single-atom state $\left|r\right>_k$. We now introduce, in addition to the single-atom ground state $\left|g\right>_k$ a second stable state $\left|s\right>_k$. The latter will in practice be constituted by a hyperfine state of the atomic ground state manifold different from $\left|g\right>_k$.

Using this additional state we can prepare the ground state (\ref{eq:ground_state}) starting from the initial state $\left|\mathrm{init}\right>$ by a sequence of two resonant laser pulses with Rabi frequencies $\Omega_{1,2}$. We choose the first pulse to be resonant on the single atom transition $\left|g\right>_k\rightarrow\left|s\right>_k$ for a time $\tau_1=\pi/(2\Omega_1)$. Subsequently we irradiate a strong laser ($\Omega_2\gg V$) that resonantly couples  $\left|s\right>_k\rightarrow\left|r\right>_k$ for a time $\tau_2=\pi/\Omega_2$. This amounts to the sequence
\begin{eqnarray}\label{eq:mapping}
 \left|\mathrm{init}\right>&=&\prod_k \left|g\right>_k\\\nonumber
 &\stackrel{\tau_1}{\rightarrow}& \prod_k \frac{1}{\sqrt{2}}\left[\left|g\right>_k+i\left|s\right>_k\right]\\\nonumber
 &\stackrel{\tau_2}{\rightarrow}& \prod_k \frac{1}{\sqrt{2}}\left[\left|g\right>_k-\left|r\right>_k\right]=\prod_k \left|-\right>_k=\left|0\right>
\end{eqnarray}
and thus results in the desired preparation of $\left|0\right>$.

\subsubsection{Excitation of many-body quantum states from $\left|0\right>$: }
Let us show now how to address the single-fermion and two-fermion states from $\left|0\right>$. To this end we make use of the Hamiltonian $H_1$ (\ref{eqn:H_1}) which emerged from the laser detuning after the application of the unitary transformation (\ref{eq:rotation}). As we discussed earlier, this term drives transitions between manifolds with $\Delta m=\pm1$, (see Fig. \ref{fig:spectrum}) and such a transition would exactly lead to the excitation of the desired many-particle states. However, we have also estimated earlier that these transitions have small probability since they are suppressed by a factor $\sim \Delta/\Omega^2$ with $|\Delta|\ll\Omega$.

\begin{figure}
\center
\includegraphics[width=7cm]{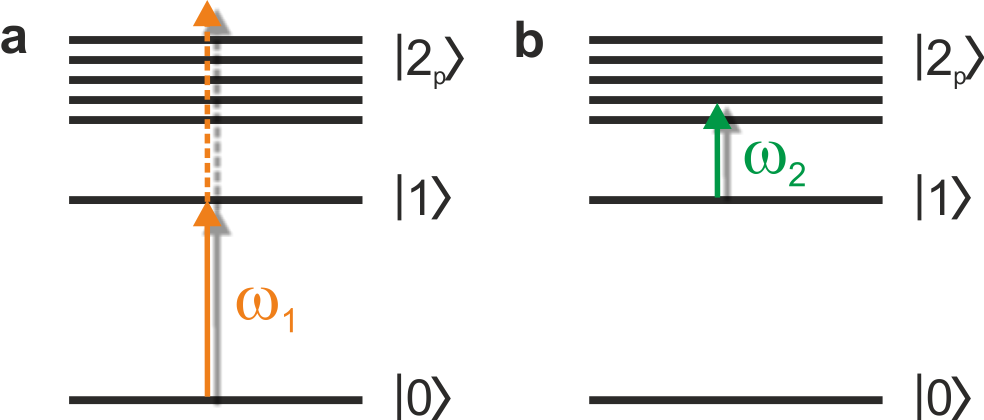}
\caption{Sketch of the excitation of the single-fermion and two-fermion states by means of an oscillating (radiofrequency) detuning using a not too large value of $\Omega$. \textbf{a:} In a first step, the population is transferred by a $\pi$-pulse to the single-fermion state by tuning the frequency of the detuning on resonance with the gap $\omega_\Delta=\omega_1=E_1-E_0$. \textbf{b:} A second $\pi$-pulse with $\omega_\Delta$ tuned to match $\omega_2=E_{2p}-E_1$ addresses the corresponding $\left|2_p\right>$ state, bearing in mind that $V\gg\Delta_\mathrm{osc}$ in this step.}\label{fig:excitation}
\end{figure}

To overcome this problem we introduce an oscillating detuning of the form $\Delta(t)=\Delta_\mathrm{osc}\cos{\left(\omega_\Delta\, t\right)}$. If one now tunes $\omega_\Delta$ such that it coincides with the gap between two given states, this detuning acts effectively as a laser that couples them resonantly (see Fig. \ref{fig:excitation}a). The matrix element corresponding to this transition (within the rotating-wave approximation) and thus the transition rate is given by
\begin{eqnarray}
  \left<\mathrm{final}\right|H_1\left|0\right>= \frac{\Delta_\mathrm{osc}}{4}\left<\mathrm{final}\right|\sum_{k=1}^L\sigma_{x}^{(k)}\left|0\right>.
\end{eqnarray}
It turns out that, due to the symmetry of the Hamiltonian and the operator $H_1$, this matrix element is non-zero only for a single final state given by
\begin{eqnarray}
  \left|1\right>=\eta_{\mathrm{o},L}^\dagger\left|0\right>=\frac{1}{\sqrt{L}}\sum_{k=1}^L\sigma_+^{(k)}\left|0\right>\label{eq:single_excitation}
\end{eqnarray}
whose energy is
\begin{eqnarray}
  E_1=E_0+2\Omega+\frac{V}{2},
\end{eqnarray}
with $E_{0}=-L\left(\Omega-\frac{V}{4}\right)$ being the ground state energy. This state is a spin wave or, in other words, a superatom that extends over the entire lattice.

Once this spin wave is excited, one can use the oscillating detuning to reach states that carry two excitations. Again, the selection rules restrict the number of accessible states and only the transitions to
\begin{eqnarray}
\left|2_p\right>&=&\eta^\dagger_{\mathrm{e},p}\eta^\dagger_{\mathrm{e},L-p+1}\left|0\right>\\
&=&\frac{2}{iL}\sum_{k>k'}\sin{\left[\frac{2\pi}{L}(p-1/2)(k-k')\right]}\sigma_+^{(k)}\sigma_+^{(k')}\left|0\right>\nonumber
\end{eqnarray}
are permitted. The energy of these states is given by
\begin{eqnarray}
  E_{2p}=E_0+4\Omega+V\cos{\left[\frac{2\pi}{L}(p-1/2)\right]}.
\end{eqnarray}
A schematics of the envisioned excitation path is depicted in Fig. \ref{fig:excitation}.
There are some limitations to bear in mind in this scheme. Firstly, in order to avoid accidental resonances during the excitation process, one can exploit the second-order level shifts that are caused by $H_2$ in the regime in which the ratio $V/\Omega$ is not too small. This is sketched in Fig. \ref{fig:excitation}a, where the gap between $\left| 1\right>$ and any of the  $\left|2_p\right>$ levels, i.e., $\omega_2=E_{2p}-E_1$, becomes increasingly different from the energy gap between $\left|0\right>$ and $\left|1\right>$, i.e., $\omega_1$, for sufficiently large $V/\Omega$. Secondly, the separation between neighboring $\left|2_p\right>$ states is of the order of $V$ and the Rabi frequency of the transition is proportional to $\Delta_\mathrm{osc}$. As a consequence, to populate only a single level of the doubly-excited manifold, the parameters have to accomplish that $V\gg\Delta_\mathrm{osc}$ and, at the same time, $\Delta_\mathrm{osc}$ has to be large enough in order to perform the transfer in a time interval that is much shorter than the lifetime of the Rydberg state.

\subsection{A single photon source}
A collective atomic excitation stored in the Rydberg ring can be converted into photons \cite{Saffman02,Porras08,Olmos10-2,Pedersen09,Nielsen10}. The particular features of the atomic state are reflected in the emission properties, i.e., the angular distribution of the emitted radiation. We will illustrate this using the single excitation state (\ref{eq:single_excitation}) as an example.

\begin{figure}
\center
\includegraphics[width=8.8cm]{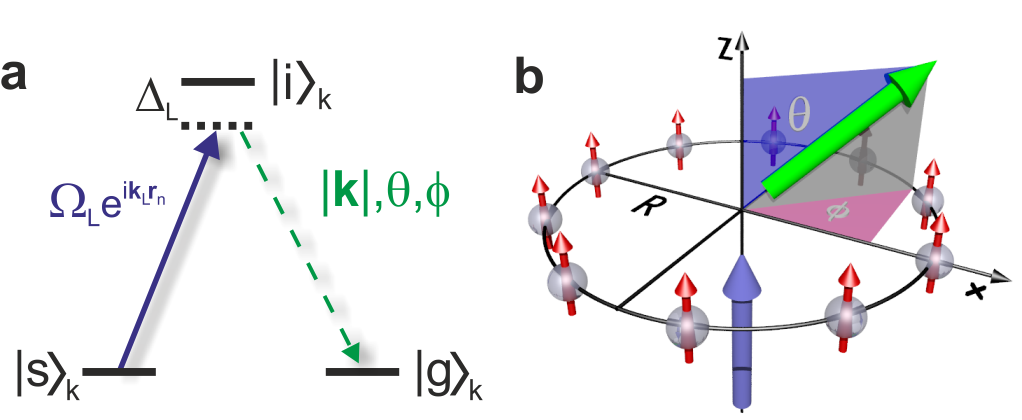}
\caption{\textbf{a:} Level scheme for the photon generation scheme. The delocalized atomic excitation is stored in the local stable states $\left|g\right>_k$ and $\left|s\right>_k$ which is achieved by a suitable mapping sequence converting states $\left|\pm\right>_k$ into ground states. The state $\left|g\right>_k$ is coupled by a laser with wavevector $\mathbf{k}_L$ and Rabi frequency $\Omega_L$ off-resonantly (detuning $\Delta_L$) to the intermediate state $\left|i\right>_k$. The photon is then emitted on the transition $\left|i\right>_k\rightarrow \left|g\right>_k$. \textbf{b:} Outcoupling scheme. We consider a situation in which the atomic transition dipole moments are parallel to the $z$-axis (small arrows) and the wavevector of the laser is $\mathbf{k}_L=|\mathbf{k}_L|\mathbf{e}_z$. Photons are then emitted into modes with a certain angular distribution, parameterized by the angles $\theta$ and $\phi$.} \label{fig:photon_creation}
\end{figure}
We start by mapping the excited state which is encoded in the superposition states $\left|+\right>_k$ and $\left|-\right>_k$ to stable states of the atomic hyperfine groundstate manifold. This is necessary since such states are much longer lived than the Rydberg states which decay due to black-body radiation or spontaneous emission. Reverting the sequence (\ref{eq:mapping}) - which led to the preparation of the ground state $\left|0\right>$ of $H_\mathrm{xy}$ - performs the mapping
\begin{eqnarray}
  \left|-\right>_k&\rightarrow&\left|g\right>_k\\
  \left|+\right>_k&\rightarrow&i\left|s\right>_k.
\end{eqnarray}
In order to convert the collective atomic excitation stored in the Rydberg ring into photons we make use of the level scheme depicted in Fig. \ref{fig:photon_creation}a \cite{Porras08}. Here a laser with wavevector $\mathbf{k}_L$ couples the state $\left|g\right>_k$ off-resonantly to an intermediate state $\left|i\right>_k$. The Rabi frequency and the detuning of this transition are given by $\Omega_L$ and $\Delta_L$, respectively. Photons are then emitted on the transition $\left|i\right>_k\rightarrow \left|g\right>_k$. In our scheme we do not consider decay from $\left|i\right>_k$ back to $\left|s\right>_k$, which can be ensured by an appropriate choice of atomic levels \cite{Porras08,Olmos10-2}.

We consider now times much longer than the lifetime of the intermediate state $\tau=\Gamma^{-1}$, where $\Gamma$ is the corresponding decay rate. One can show that in the above described scheme a single atomic excitation is mapped to a single photon state according to
\begin{eqnarray}
  \sigma^+_j \left|0\right>\rightarrow \sum_{\mathbf{q},\nu} g_{j\mathbf{q}\nu} a^\dagger_{\mathbf{q}\nu}\left|\mathrm{vac}\right>
\end{eqnarray}
where $a^\dagger_{\mathbf{q}\nu}$ creates a photon with momentum $\mathbf{q}$ and polarization $\nu$ and $\left|\mathrm{vac}\right>$ is the photon vacuum.
The coefficients $g_{j\mathbf{q}\nu} $ contain information about the coupling strength of the excitation on the $j$-th lattice site to the photon mode characterized by $\mathbf{q}$ and $\nu$, and are given by
\begin{equation}\label{eq:g}
  g_{j \mathbf{q}\nu}(t)=-iK_{\mathbf{q}\nu}e^{-i\omega t}\sum_{\gamma k}e^{-i\mathbf{q}\cdot\mathbf{r}_\gamma}\frac{{\cal M}_{\gamma k}{\cal M}^{-1}_{kj}}{i\omega-D_k}.
\end{equation}
Here, $\omega=|\mathbf{q}|/c$, $\mathbf{r}_\gamma\equiv R\left(\cos{\phi_\gamma},\sin{\phi_\gamma},0\right)$ with $\phi_{\gamma}=\frac{2\pi}{L}(\gamma-1)$ denotes the position of the atoms and the coefficient $K_{\mathbf{q}\nu}$ reads
\begin{equation}\label{eqn:Kq}
  K_{\mathbf{q}\nu}=\left(\frac{\Omega_L}{\Delta_L}\right)\sqrt{\frac{\omega}{2\epsilon_0 V}}\mathbf{d_{\mathrm{gi}}}\cdot\mathbf{e}_{\mathbf{q}\nu},
\end{equation}
where $V$ is the quantization volume, $\epsilon_0$ the vacuum permitivity, and $\mathbf{d_{\mathrm{gi}}}$ is the dipole operator of the $\left|g\right>_k\rightarrow\left|i\right>_k$ transition. $\mathbf{{\cal M}}_k$ and $D_k$ are the eigenvectors and eigenvalues of the operator that governs the atomic dynamics and which we call $J$. It depends on the orientation of the atomic transition dipole moments and the quantity $k_LR$, where $R\approx aL/ 2\pi$ is the radius of the ring. We consider the particularly simple situation in which the transition dipole moments are aligned with the $z$-axis (see Fig. \ref{fig:photon_creation}b). In this case, the matrix representation of $J$ is a circulant complex symmetric matrix \cite{Tee05} and its eigenvalues and eigenfunctions are given by
\begin{equation}\label{eq:eigen}
  {\cal M}_{\gamma k}=\frac{e^{i\frac{2\pi}{L}(\gamma-1)(k-1)}}{\sqrt{L}}\qquad D_k=\sum_{n=1}^{L}J_{1n}e^{i\frac{2\pi}{L}(n-1)(k-1)},
\end{equation}
where $J_{1n}=\gamma_{1n}+i\Omega_{1n}$ and
\begin{eqnarray*}
  \gamma_{1n}&=&\frac{3\Gamma}{2}\left[\frac{\cos{\kappa_{1n}}}{\kappa_{1n}^2}- \frac{\sin{\kappa_{1n}}}{\kappa_{1n}^3}+\frac{\sin{\kappa_{1n}}}{\kappa_{1n}}\right]\\
  \Omega_{1n}&=&\frac{3\Gamma}{2}\left[\frac{\sin{\kappa_{1n}}}{\kappa_{1n}^2}+ \frac{\cos{\kappa_{1n}}}{\kappa_{1n}^3}-\frac{\cos{\kappa_{1n}}}{\kappa_{1n}}\right],
\end{eqnarray*}
with $\kappa_{1n}=k_LR|\hat{\mathbf{r}}_{1}-\hat{\mathbf{r}}_{n}|$ \cite{Lehmberg70}.

To study the angular distribution of the emitted photon we study the intensity of photons per solid angle that is defined through
\begin{eqnarray}
  I(\theta,\phi)=\frac{V}{(2\pi c)^3}\int_0^\infty \sum_{\nu}\left<n_{\mathbf{q}\nu}\right>\omega^2d\omega.
\end{eqnarray}
In the particular case of the spin wave given in (\ref{eq:single_excitation}) and the direction of $\mathbf{k}_L$ being parallel to the $z$-axis, one can prove that the expression of the angular intensity yields
\begin{equation}
  I(\theta,\phi)=\frac{3\Gamma}{4\pi L}\frac{\sin^2{\theta}}{D_1+D_1^*}\left|\sum_{\gamma=1}^Le^{-ik_LR\, \hat{\mathbf{q}}\cdot\hat{\mathbf{r}_{\gamma}}}\right|^2,
\end{equation}
with $\hat{\mathbf{q}}=(\sin{\theta}\cos{\phi},\sin{\theta}\sin{\phi},\cos{\theta})$. For sufficiently small values of $k_La$, i.e., $k_La\lesssim\pi/2$, the sum in the previous expression can be substituted by a Bessel function and the intensity becomes
\begin{equation}\label{eq:Int-simplified}
  I(\theta,\phi)\approx\frac{3\Gamma L}{4\pi}\frac{\sin^2{\theta}}{D_1+D_1^*}J_0^2(k_LR\sin{\theta}).
\end{equation}
\begin{figure}
\center
\includegraphics[width=8.8cm]{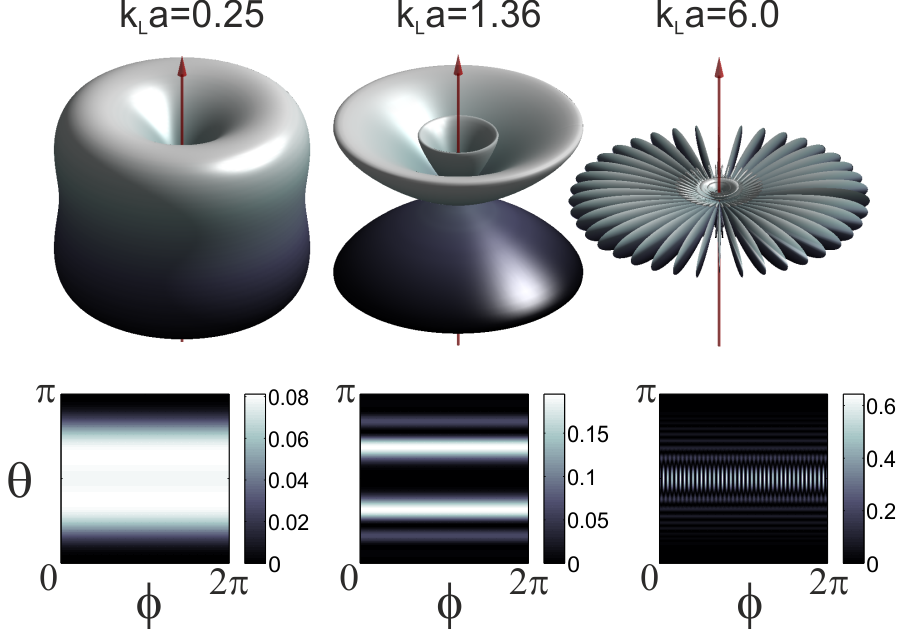}
\caption{Angular intensity distribution for a ring of $L$ sites and various values of $k_La$. The density plots in the lower row show the same data as the three-dimensional plots. The red arrow indicates the direction of $\mathbf{k}_L$ which is chosen to be parallel to the $z$-axis. For $k_La\lesssim\pi/2$ the intensity profile is created through the collective coupling of the atoms to the radiation field. Here no dependence on the azimuthal angle is visible. For larger $k_La$ the atoms couple individually to the electromagnetic field which gives rise to a large number of peaks.} \label{fig:intensity}
\end{figure}
In Fig. \ref{fig:intensity} we show the intensity for $k_La=(0.25,\,1.36,\,6.0)$ and $L=40$. The figure shows strikingly how the distance between the atoms affects the profile of the radiation. For $k_La=0.25$ the atoms are so close together that the atomic excitation (spin wave) acts as a single degree of freedom that couples to the radiation field \cite{Lehmberg70}. This results in an almost spherical intensity profile which is modulated by the dipole radiation pattern. In the intermediate case $k_La=1.36$ the coupling to the radiation field is still dominated by collective effects. Here, the photon emission is strongly peaked along a polar angle $\theta\approx \pi/3$. The position of this peak can be controlled by tuning the parameter $k_La$. In the previous two cases, the intensity profiles show no dependence on the azimuthal angle, as we expected from expression (\ref{eq:Int-simplified}). In the third case, $k_La=6.0$, the equation (\ref{eq:Int-simplified}) is no longer valid, and the atoms can be approximately regarded as independent so that they are coupled individually to the radiation field. This produces a large number $L$ of peaks, as a result of the interference of the $L$ sites.

In addition to these examples the Rydberg ring also allows to generate single photons with a strong directionality, i.e. the photon emission takes place only into a very small solid angle. Moreover, by using many-body states that contain more than a single excitation, the creation of correlated photon pairs or triples can be achieved \cite{Olmos10-2}.

\section{Conclusions and Outlook}
In this paper we could merely offer a slight glimpse on the potential the highly excited atoms have to offer for the study and the understanding of quantum many-body physics. We saw that already the very simple Rydberg ring allows to tackle such fundamental questions as the thermalization of a closed quantum systems. Moreover, we demonstrated that the Rydberg ring can be used to create single photon sources with particular emission characteristics.

More interesting features are expected to emerge in higher dimensional setups, particularly in two dimensions where many-body systems possess a particularly rich phase structure. While certainly some of the concepts that were presented here are also applicable in higher dimensions there will also be fundamental differences. E.g., in the limit of strong laser driving the system will no longer be analytically solvable as the Jordan-Wigner transformation leads only to a solvable model in one dimension.

The current experimental situation is such that Rydberg atoms are excited from an atomic gas in which the atoms are randomly distributed, and the experimental realization of the Rydberg ring is certainly several steps further down the road. Significant efforts are, however, undertaken to study the Rydberg excitation dynamics in structured environments and first exciting results have shown the feasibility of these undertakings \cite{Urban08,Gaetan08,Isenhower10,Wilk10}.

The authors acknowledge funding through EPSRC.

\footnotesize{
\providecommand*{\mcitethebibliography}{\thebibliography}
\csname @ifundefined\endcsname{endmcitethebibliography}
{\let\endmcitethebibliography\endthebibliography}{}

\bibliographystyle{rsc} 

\begin{mcitethebibliography}{56}
\providecommand*{\natexlab}[1]{#1}
\providecommand*{\mciteSetBstSublistMode}[1]{}
\providecommand*{\mciteSetBstMaxWidthForm}[2]{}
\providecommand*{\mciteBstWouldAddEndPuncttrue}
  {\def\EndOfBibitem{\unskip.}}
\providecommand*{\mciteBstWouldAddEndPunctfalse}
  {\let\EndOfBibitem\relax}
\providecommand*{\mciteSetBstMidEndSepPunct}[3]{}
\providecommand*{\mciteSetBstSublistLabelBeginEnd}[3]{}
\providecommand*{\EndOfBibitem}{}
\mciteSetBstSublistMode{f}
\mciteSetBstMaxWidthForm{subitem}
{(\emph{\alph{mcitesubitemcount}})}
\mciteSetBstSublistLabelBeginEnd{\mcitemaxwidthsubitemform\space}
{\relax}{\relax}

\bibitem[Metcalf and Straten(1999)]{Metcalf99}
H.~J. Metcalf and P.~v.~d. Straten, \emph{Laser cooling and trapping}, Springer
  Verlag-New York, 1999\relax
\mciteBstWouldAddEndPuncttrue
\mciteSetBstMidEndSepPunct{\mcitedefaultmidpunct}
{\mcitedefaultendpunct}{\mcitedefaultseppunct}\relax
\EndOfBibitem
\bibitem[Bloch \emph{et~al.}(2008)Bloch, Dalibard, and Zwerger]{Bloch08}
I.~Bloch, J.~Dalibard and W.~Zwerger, \emph{Rev. Mod. Phys.}, 2008,
  \textbf{80}, 885\relax
\mciteBstWouldAddEndPuncttrue
\mciteSetBstMidEndSepPunct{\mcitedefaultmidpunct}
{\mcitedefaultendpunct}{\mcitedefaultseppunct}\relax
\EndOfBibitem
\bibitem[Greiner \emph{et~al.}(2002)Greiner, Mandel, Esslinger, H\"ansch, and
  Bloch]{Greiner02}
M.~Greiner, O.~Mandel, T.~Esslinger, T.~W. H\"ansch and I.~Bloch,
  \emph{Nature}, 2002, \textbf{415}, 39--44\relax
\mciteBstWouldAddEndPuncttrue
\mciteSetBstMidEndSepPunct{\mcitedefaultmidpunct}
{\mcitedefaultendpunct}{\mcitedefaultseppunct}\relax
\EndOfBibitem
\bibitem[Hadzibabic \emph{et~al.}(2006)Hadzibabic, Kr\"uger, Cheneau,
  Battelier, and Dalibard]{Hadzibabic06}
Z.~Hadzibabic, P.~Kr\"uger, M.~Cheneau, B.~Battelier and J.~Dalibard,
  \emph{Nature}, 2006, \textbf{441}, 1118--1121\relax
\mciteBstWouldAddEndPuncttrue
\mciteSetBstMidEndSepPunct{\mcitedefaultmidpunct}
{\mcitedefaultendpunct}{\mcitedefaultseppunct}\relax
\EndOfBibitem
\bibitem[Haller \emph{et~al.}(2010)Haller, Hart, Mark, Danzl, Reichsollner,
  Gustavsson, Dalmonte, Pupillo, and N\"agerl]{Haller10}
E.~Haller, R.~Hart, M.~J. Mark, J.~G. Danzl, L.~Reichsollner, M.~Gustavsson,
  M.~Dalmonte, G.~Pupillo and H.-C. N\"agerl, \emph{Nature}, 2010,
  \textbf{466}, 597–600\relax
\mciteBstWouldAddEndPuncttrue
\mciteSetBstMidEndSepPunct{\mcitedefaultmidpunct}
{\mcitedefaultendpunct}{\mcitedefaultseppunct}\relax
\EndOfBibitem
\bibitem[Gallagher(1984)]{Gallagher84}
T.~Gallagher, \emph{Rydberg Atoms}, Cambridge University Press, 1984\relax
\mciteBstWouldAddEndPuncttrue
\mciteSetBstMidEndSepPunct{\mcitedefaultmidpunct}
{\mcitedefaultendpunct}{\mcitedefaultseppunct}\relax
\EndOfBibitem
\bibitem[Marinescu and Dalgarno(1995)]{Marinescu95}
M.~Marinescu and A.~Dalgarno, \emph{Phys. Rev. A}, 1995, \textbf{52},
  311--328\relax
\mciteBstWouldAddEndPuncttrue
\mciteSetBstMidEndSepPunct{\mcitedefaultmidpunct}
{\mcitedefaultendpunct}{\mcitedefaultseppunct}\relax
\EndOfBibitem
\bibitem[Mourachko \emph{et~al.}(1998)Mourachko, Comparat, de~Tomasi, Fioretti,
  Nosbaum, Akulin, and Pillet]{Mourachko98}
I.~Mourachko, D.~Comparat, F.~de~Tomasi, A.~Fioretti, P.~Nosbaum, V.~M. Akulin
  and P.~Pillet, \emph{Phys. Rev. Lett.}, 1998, \textbf{80}, 253--256\relax
\mciteBstWouldAddEndPuncttrue
\mciteSetBstMidEndSepPunct{\mcitedefaultmidpunct}
{\mcitedefaultendpunct}{\mcitedefaultseppunct}\relax
\EndOfBibitem
\bibitem[Anderson \emph{et~al.}(1998)Anderson, Veale, and
  Gallagher]{Anderson98}
W.~R. Anderson, J.~Veale and T.~F. Gallagher, \emph{Phys. Rev. Lett.}, 1998,
  \textbf{80}, 249--252\relax
\mciteBstWouldAddEndPuncttrue
\mciteSetBstMidEndSepPunct{\mcitedefaultmidpunct}
{\mcitedefaultendpunct}{\mcitedefaultseppunct}\relax
\EndOfBibitem
\bibitem[Tong \emph{et~al.}(2004)Tong, Farooqi, Stanojevic, Krishnan, Zhang,
  C\^ot\'e, Eyler, and Gould]{Tong04}
D.~Tong, S.~M. Farooqi, J.~Stanojevic, S.~Krishnan, Y.~P. Zhang, R.~C\^ot\'e,
  E.~E. Eyler and P.~L. Gould, \emph{Phys. Rev. Lett.}, 2004, \textbf{93},
  063001\relax
\mciteBstWouldAddEndPuncttrue
\mciteSetBstMidEndSepPunct{\mcitedefaultmidpunct}
{\mcitedefaultendpunct}{\mcitedefaultseppunct}\relax
\EndOfBibitem
\bibitem[Singer \emph{et~al.}(2004)Singer, Reetz-Lamour, Amthor, Marcassa, and
  Weidem\"uller]{Singer04}
K.~Singer, M.~Reetz-Lamour, T.~Amthor, L.~G. Marcassa and M.~Weidem\"uller,
  \emph{Phys. Rev. Lett.}, 2004, \textbf{93}, 163001\relax
\mciteBstWouldAddEndPuncttrue
\mciteSetBstMidEndSepPunct{\mcitedefaultmidpunct}
{\mcitedefaultendpunct}{\mcitedefaultseppunct}\relax
\EndOfBibitem
\bibitem[Lukin \emph{et~al.}(2001)Lukin, Fleischhauer, C\^{o}t\'{e}, Duan,
  Jaksch, Cirac, and Zoller]{Lukin01}
M.~D. Lukin, M.~Fleischhauer, R.~C\^{o}t\'{e}, L.~M. Duan, D.~Jaksch, J.~I.
  Cirac and P.~Zoller, \emph{Phys. Rev. Lett.}, 2001, \textbf{87}, 037901\relax
\mciteBstWouldAddEndPuncttrue
\mciteSetBstMidEndSepPunct{\mcitedefaultmidpunct}
{\mcitedefaultendpunct}{\mcitedefaultseppunct}\relax
\EndOfBibitem
\bibitem[Jaksch \emph{et~al.}(2000)Jaksch, Cirac, Zoller, Rolston,
  C\^{o}t\'{e}, and Lukin]{Jaksch00}
D.~Jaksch, J.~I. Cirac, P.~Zoller, S.~L. Rolston, R.~C\^{o}t\'{e} and M.~D.
  Lukin, \emph{Phys. Rev. Lett.}, 2000, \textbf{85}, 2208--2211\relax
\mciteBstWouldAddEndPuncttrue
\mciteSetBstMidEndSepPunct{\mcitedefaultmidpunct}
{\mcitedefaultendpunct}{\mcitedefaultseppunct}\relax
\EndOfBibitem
\bibitem[Liebisch \emph{et~al.}(2005)Liebisch, Reinhard, Berman, and
  Raithel]{CubelLiebisch05}
T.~C. Liebisch, A.~Reinhard, P.~R. Berman and G.~Raithel, \emph{Phys. Rev.
  Lett.}, 2005, \textbf{95}, 253002\relax
\mciteBstWouldAddEndPuncttrue
\mciteSetBstMidEndSepPunct{\mcitedefaultmidpunct}
{\mcitedefaultendpunct}{\mcitedefaultseppunct}\relax
\EndOfBibitem
\bibitem[Heidemann \emph{et~al.}(2007)Heidemann, Raitzsch, Bendkowsky,
  Butscher, L\"{o}w, Santos, and Pfau]{Heidemann07}
R.~Heidemann, U.~Raitzsch, V.~Bendkowsky, B.~Butscher, R.~L\"{o}w, L.~Santos
  and T.~Pfau, \emph{Phys. Rev. Lett.}, 2007, \textbf{99}, 163601\relax
\mciteBstWouldAddEndPuncttrue
\mciteSetBstMidEndSepPunct{\mcitedefaultmidpunct}
{\mcitedefaultendpunct}{\mcitedefaultseppunct}\relax
\EndOfBibitem
\bibitem[Reetz-Lamour \emph{et~al.}(2008)Reetz-Lamour, Amthor, Deiglmayr, and
  Weidem\"{u}ller]{Reetz-Lamour08}
M.~Reetz-Lamour, T.~Amthor, J.~Deiglmayr and M.~Weidem\"{u}ller, \emph{Phys.
  Rev. Lett.}, 2008, \textbf{100}, 253001\relax
\mciteBstWouldAddEndPuncttrue
\mciteSetBstMidEndSepPunct{\mcitedefaultmidpunct}
{\mcitedefaultendpunct}{\mcitedefaultseppunct}\relax
\EndOfBibitem
\bibitem[Pritchard \emph{et~al.}(2010)Pritchard, Maxwell, Gauguet, Weatherill,
  Jones, and Adams]{Pritchard10}
J.~D. Pritchard, D.~Maxwell, A.~Gauguet, K.~J. Weatherill, M.~P.~A. Jones and
  C.~S. Adams, \emph{arXiv:1006.4087}, 2010\relax
\mciteBstWouldAddEndPuncttrue
\mciteSetBstMidEndSepPunct{\mcitedefaultmidpunct}
{\mcitedefaultendpunct}{\mcitedefaultseppunct}\relax
\EndOfBibitem
\bibitem[Urban \emph{et~al.}(2009)Urban, Johnson, Henage, Isenhower, Yavuz,
  Walker, and Saffman]{Urban08}
E.~Urban, T.~A. Johnson, T.~Henage, L.~Isenhower, D.~D. Yavuz, T.~G. Walker and
  M.~Saffman, \emph{Nature Phys.}, 2009, \textbf{5}, 110\relax
\mciteBstWouldAddEndPuncttrue
\mciteSetBstMidEndSepPunct{\mcitedefaultmidpunct}
{\mcitedefaultendpunct}{\mcitedefaultseppunct}\relax
\EndOfBibitem
\bibitem[Ga\"{e}tan \emph{et~al.}(2009)Ga\"{e}tan, Miroshnychenko, Wilk,
  Chotia, Viteau, Comparat, Pillet, Browaeys, and Grangier]{Gaetan08}
A.~Ga\"{e}tan, Y.~Miroshnychenko, T.~Wilk, A.~Chotia, M.~Viteau, D.~Comparat,
  P.~Pillet, A.~Browaeys and P.~Grangier, \emph{Nature Phys.}, 2009,
  \textbf{5}, 115\relax
\mciteBstWouldAddEndPuncttrue
\mciteSetBstMidEndSepPunct{\mcitedefaultmidpunct}
{\mcitedefaultendpunct}{\mcitedefaultseppunct}\relax
\EndOfBibitem
\bibitem[Isenhower \emph{et~al.}(2010)Isenhower, Urban, Zhang, Gill, Henage,
  Johnson, Walker, and Saffman]{Isenhower10}
L.~Isenhower, E.~Urban, X.~L. Zhang, A.~T. Gill, T.~Henage, T.~A. Johnson,
  T.~G. Walker and M.~Saffman, \emph{Phys. Rev. Lett.}, 2010, \textbf{104},
  010503\relax
\mciteBstWouldAddEndPuncttrue
\mciteSetBstMidEndSepPunct{\mcitedefaultmidpunct}
{\mcitedefaultendpunct}{\mcitedefaultseppunct}\relax
\EndOfBibitem
\bibitem[Wilk \emph{et~al.}(2010)Wilk, Ga\"etan, Evellin, Wolters,
  Miroshnychenko, Grangier, and Browaeys]{Wilk10}
T.~Wilk, A.~Ga\"etan, C.~Evellin, J.~Wolters, Y.~Miroshnychenko, P.~Grangier
  and A.~Browaeys, \emph{Phys. Rev. Lett.}, 2010, \textbf{104}, 010502\relax
\mciteBstWouldAddEndPuncttrue
\mciteSetBstMidEndSepPunct{\mcitedefaultmidpunct}
{\mcitedefaultendpunct}{\mcitedefaultseppunct}\relax
\EndOfBibitem
\bibitem[Saffman \emph{et~al.}(2009)Saffman, Walker, and
  M{\o}lmer]{Saffman09-2}
M.~Saffman, T.~G. Walker and K.~M{\o}lmer, \emph{arXiv:0909.4777v1}, 2009\relax
\mciteBstWouldAddEndPuncttrue
\mciteSetBstMidEndSepPunct{\mcitedefaultmidpunct}
{\mcitedefaultendpunct}{\mcitedefaultseppunct}\relax
\EndOfBibitem
\bibitem[Weimer \emph{et~al.}(2008)Weimer, L\"{o}w, Pfau, and
  B\"{u}chler]{Weimer08}
H.~Weimer, R.~L\"{o}w, T.~Pfau and H.~P. B\"{u}chler, \emph{Phys. Rev. Lett.},
  2008, \textbf{101}, 250601\relax
\mciteBstWouldAddEndPuncttrue
\mciteSetBstMidEndSepPunct{\mcitedefaultmidpunct}
{\mcitedefaultendpunct}{\mcitedefaultseppunct}\relax
\EndOfBibitem
\bibitem[Olmos \emph{et~al.}(2010)Olmos, M\"{u}ller, and Lesanovsky]{Olmos09-2}
B.~Olmos, M.~M\"{u}ller and I.~Lesanovsky, \emph{New Journal of Physics}, 2010,
  \textbf{12}, 013024\relax
\mciteBstWouldAddEndPuncttrue
\mciteSetBstMidEndSepPunct{\mcitedefaultmidpunct}
{\mcitedefaultendpunct}{\mcitedefaultseppunct}\relax
\EndOfBibitem
\bibitem[Lesanovsky \emph{et~al.}(2010)Lesanovsky, Olmos, and
  Garrahan]{Lesanovsky10}
I.~Lesanovsky, B.~Olmos and J.~Garrahan, \emph{arXiv:1004.3210, to appear in
  Phys. Rev. Lett.}, 2010\relax
\mciteBstWouldAddEndPuncttrue
\mciteSetBstMidEndSepPunct{\mcitedefaultmidpunct}
{\mcitedefaultendpunct}{\mcitedefaultseppunct}\relax
\EndOfBibitem
\bibitem[Olmos \emph{et~al.}(2009)Olmos, Gonz\'{a}lez-F\'{e}rez, and
  Lesanovsky]{Olmos09-3}
B.~Olmos, R.~Gonz\'{a}lez-F\'{e}rez and I.~Lesanovsky, \emph{Phys. Rev. Lett.},
  2009, \textbf{103}, 185302\relax
\mciteBstWouldAddEndPuncttrue
\mciteSetBstMidEndSepPunct{\mcitedefaultmidpunct}
{\mcitedefaultendpunct}{\mcitedefaultseppunct}\relax
\EndOfBibitem
\bibitem[Pohl \emph{et~al.}(2010)Pohl, Demler, and Lukin]{Pohl10}
T.~Pohl, E.~Demler and M.~D. Lukin, \emph{Phys. Rev. Lett.}, 2010,
  \textbf{104}, 043002\relax
\mciteBstWouldAddEndPuncttrue
\mciteSetBstMidEndSepPunct{\mcitedefaultmidpunct}
{\mcitedefaultendpunct}{\mcitedefaultseppunct}\relax
\EndOfBibitem
\bibitem[Schachenmayer \emph{et~al.}(2010)Schachenmayer, Lesanovsky, and
  Daley]{Schachenmayer10}
J.~Schachenmayer, I.~Lesanovsky and A.~Daley, \emph{in preparation}, 2010\relax
\mciteBstWouldAddEndPuncttrue
\mciteSetBstMidEndSepPunct{\mcitedefaultmidpunct}
{\mcitedefaultendpunct}{\mcitedefaultseppunct}\relax
\EndOfBibitem
\bibitem[Weimer \emph{et~al.}(2010)Weimer, M\"{u}ller, Lesanovsky, Zoller, and
  B\"{u}chler]{Weimer10}
H.~Weimer, M.~M\"{u}ller, I.~Lesanovsky, P.~Zoller and H.~P. B\"{u}chler,
  \emph{Nature Physics}, 2010, \textbf{6}, 382--388\relax
\mciteBstWouldAddEndPuncttrue
\mciteSetBstMidEndSepPunct{\mcitedefaultmidpunct}
{\mcitedefaultendpunct}{\mcitedefaultseppunct}\relax
\EndOfBibitem
\bibitem[Pupillo \emph{et~al.}(2010)Pupillo, Micheli, Boninsegni, Lesanovsky,
  and Zoller]{Pupillo10}
G.~Pupillo, A.~Micheli, M.~Boninsegni, I.~Lesanovsky and P.~Zoller, \emph{Phys.
  Rev. Lett.}, 2010, \textbf{104}, 223002\relax
\mciteBstWouldAddEndPuncttrue
\mciteSetBstMidEndSepPunct{\mcitedefaultmidpunct}
{\mcitedefaultendpunct}{\mcitedefaultseppunct}\relax
\EndOfBibitem
\bibitem[Mayle \emph{et~al.}(2009)Mayle, Lesanovsky, and Schmelcher]{Mayle09}
M.~Mayle, I.~Lesanovsky and P.~Schmelcher, \emph{Phys. Rev. A}, 2009,
  \textbf{79}, 041403\relax
\mciteBstWouldAddEndPuncttrue
\mciteSetBstMidEndSepPunct{\mcitedefaultmidpunct}
{\mcitedefaultendpunct}{\mcitedefaultseppunct}\relax
\EndOfBibitem
\bibitem[Henkel \emph{et~al.}(2010)Henkel, Nath, and Pohl]{Henkel10}
N.~Henkel, R.~Nath and T.~Pohl, \emph{Phys. Rev. Lett.}, 2010, \textbf{104},
  195302\relax
\mciteBstWouldAddEndPuncttrue
\mciteSetBstMidEndSepPunct{\mcitedefaultmidpunct}
{\mcitedefaultendpunct}{\mcitedefaultseppunct}\relax
\EndOfBibitem
\bibitem[Kinoshita \emph{et~al.}(2006)Kinoshita, Wenger, and
  Weiss]{Kinoshita06}
T.~Kinoshita, T.~R. Wenger and D.~S. Weiss, \emph{Nature}, 2006, \textbf{440},
  900\relax
\mciteBstWouldAddEndPuncttrue
\mciteSetBstMidEndSepPunct{\mcitedefaultmidpunct}
{\mcitedefaultendpunct}{\mcitedefaultseppunct}\relax
\EndOfBibitem
\bibitem[Hofferberth \emph{et~al.}(2007)Hofferberth, Lesanovsky, Fischer,
  Schumm, and Schmiedmayer]{Hofferberth07}
S.~Hofferberth, I.~Lesanovsky, B.~Fischer, T.~Schumm and J.~Schmiedmayer,
  \emph{Nature}, 2007, \textbf{449}, 324--327\relax
\mciteBstWouldAddEndPuncttrue
\mciteSetBstMidEndSepPunct{\mcitedefaultmidpunct}
{\mcitedefaultendpunct}{\mcitedefaultseppunct}\relax
\EndOfBibitem
\bibitem[Rigol \emph{et~al.}(2008)Rigol, Dunjko, and Olshanii]{Rigol08}
M.~Rigol, V.~Dunjko and M.~Olshanii, \emph{Nature}, 2008, \textbf{452},
  854--858\relax
\mciteBstWouldAddEndPuncttrue
\mciteSetBstMidEndSepPunct{\mcitedefaultmidpunct}
{\mcitedefaultendpunct}{\mcitedefaultseppunct}\relax
\EndOfBibitem
\bibitem[Biroli \emph{et~al.}(2009)Biroli, Kollath, and Laeuchli]{Biroli09}
G.~Biroli, C.~Kollath and A.~Laeuchli, \emph{preprint}, 2009,
  arXiv:0907.3731\relax
\mciteBstWouldAddEndPuncttrue
\mciteSetBstMidEndSepPunct{\mcitedefaultmidpunct}
{\mcitedefaultendpunct}{\mcitedefaultseppunct}\relax
\EndOfBibitem
\bibitem[Pal and Huse(2010)]{Pal10}
A.~Pal and D.~Huse, \emph{preprint}, 2010,  arXiv:1003.2613\relax
\mciteBstWouldAddEndPuncttrue
\mciteSetBstMidEndSepPunct{\mcitedefaultmidpunct}
{\mcitedefaultendpunct}{\mcitedefaultseppunct}\relax
\EndOfBibitem
\bibitem[Canovi \emph{et~al.}(2010)Canovi, Davide, Fazio, Santoro, and
  Silva]{Canovi10}
E.~Canovi, R.~Davide, R.~Fazio, G.~E. Santoro and A.~Silva, \emph{preprint},
  2010,  arXiv:1006.1634\relax
\mciteBstWouldAddEndPuncttrue
\mciteSetBstMidEndSepPunct{\mcitedefaultmidpunct}
{\mcitedefaultendpunct}{\mcitedefaultseppunct}\relax
\EndOfBibitem
\bibitem[Olmos \emph{et~al.}(2010)Olmos, Gonz\'alez-F\'erez, and
  Lesanovsky]{Olmos10-1}
B.~Olmos, R.~Gonz\'alez-F\'erez and I.~Lesanovsky, \emph{Phys. Rev. A}, 2010,
  \textbf{81}, 023604\relax
\mciteBstWouldAddEndPuncttrue
\mciteSetBstMidEndSepPunct{\mcitedefaultmidpunct}
{\mcitedefaultendpunct}{\mcitedefaultseppunct}\relax
\EndOfBibitem
\bibitem[Saffman and Walker(2002)]{Saffman02}
M.~Saffman and T.~G. Walker, \emph{Phys. Rev. A}, 2002, \textbf{66},
  065403\relax
\mciteBstWouldAddEndPuncttrue
\mciteSetBstMidEndSepPunct{\mcitedefaultmidpunct}
{\mcitedefaultendpunct}{\mcitedefaultseppunct}\relax
\EndOfBibitem
\bibitem[Porras and Cirac(2008)]{Porras08}
D.~Porras and J.~I. Cirac, \emph{Phys. Rev. A}, 2008, \textbf{78}, 053816\relax
\mciteBstWouldAddEndPuncttrue
\mciteSetBstMidEndSepPunct{\mcitedefaultmidpunct}
{\mcitedefaultendpunct}{\mcitedefaultseppunct}\relax
\EndOfBibitem
\bibitem[Olmos and Lesanovsky(2010)]{Olmos10-2}
B.~Olmos and I.~Lesanovsky, \emph{in preparation}, 2010\relax
\mciteBstWouldAddEndPuncttrue
\mciteSetBstMidEndSepPunct{\mcitedefaultmidpunct}
{\mcitedefaultendpunct}{\mcitedefaultseppunct}\relax
\EndOfBibitem
\bibitem[Pedersen and M{\o}lmer(2009)]{Pedersen09}
L.~H. Pedersen and K.~M{\o}lmer, \emph{Phys. Rev. A}, 2009, \textbf{79},
  012320\relax
\mciteBstWouldAddEndPuncttrue
\mciteSetBstMidEndSepPunct{\mcitedefaultmidpunct}
{\mcitedefaultendpunct}{\mcitedefaultseppunct}\relax
\EndOfBibitem
\bibitem[Nielsen and M\o{}lmer(2010)]{Nielsen10}
A.~E.~B. Nielsen and K.~M\o{}lmer, \emph{Phys. Rev. A}, 2010, \textbf{81},
  043822\relax
\mciteBstWouldAddEndPuncttrue
\mciteSetBstMidEndSepPunct{\mcitedefaultmidpunct}
{\mcitedefaultendpunct}{\mcitedefaultseppunct}\relax
\EndOfBibitem
\bibitem[Nelson \emph{et~al.}(2007)Nelson, Li, and Weiss]{Nelson07}
K.~D. Nelson, X.~Li and D.~S. Weiss, \emph{Nature Phys.}, 2007, \textbf{3}, 556
  -- 560\relax
\mciteBstWouldAddEndPuncttrue
\mciteSetBstMidEndSepPunct{\mcitedefaultmidpunct}
{\mcitedefaultendpunct}{\mcitedefaultseppunct}\relax
\EndOfBibitem
\bibitem[Hezel \emph{et~al.}(2006)Hezel, Lesanovsky, and Schmelcher]{Hezel06}
B.~Hezel, I.~Lesanovsky and P.~Schmelcher, \emph{Phys. Rev. Lett.}, 2006,
  \textbf{97}, 223001\relax
\mciteBstWouldAddEndPuncttrue
\mciteSetBstMidEndSepPunct{\mcitedefaultmidpunct}
{\mcitedefaultendpunct}{\mcitedefaultseppunct}\relax
\EndOfBibitem
\bibitem[Whitlock \emph{et~al.}(2009)Whitlock, Gerritsma, Fernholz, and
  Spreeuw]{whitlock09}
S.~Whitlock, R.~Gerritsma, T.~Fernholz and R.~J.~C. Spreeuw, \emph{New J.
  Phys.}, 2009, \textbf{11}, 023021\relax
\mciteBstWouldAddEndPuncttrue
\mciteSetBstMidEndSepPunct{\mcitedefaultmidpunct}
{\mcitedefaultendpunct}{\mcitedefaultseppunct}\relax
\EndOfBibitem
\bibitem[Singer \emph{et~al.}(2005)Singer, Stanojevic, Weidem\"{u}ller, and
  C\^{o}t\'{e}]{Singer05}
K.~Singer, J.~Stanojevic, M.~Weidem\"{u}ller and R.~C\^{o}t\'{e}, \emph{J.
  Phys. B: At. Mol. Opt. Phys.}, 2005, \textbf{38}, S295\relax
\mciteBstWouldAddEndPuncttrue
\mciteSetBstMidEndSepPunct{\mcitedefaultmidpunct}
{\mcitedefaultendpunct}{\mcitedefaultseppunct}\relax
\EndOfBibitem
\bibitem[Heidemann \emph{et~al.}(2008)Heidemann, Raitzsch, Bendkowsky,
  Butscher, L\"{o}w, and Pfau]{Heidemann08}
R.~Heidemann, U.~Raitzsch, V.~Bendkowsky, B.~Butscher, R.~L\"{o}w and T.~Pfau,
  \emph{Phys. Rev. Lett.}, 2008, \textbf{100}, 033601\relax
\mciteBstWouldAddEndPuncttrue
\mciteSetBstMidEndSepPunct{\mcitedefaultmidpunct}
{\mcitedefaultendpunct}{\mcitedefaultseppunct}\relax
\EndOfBibitem
\bibitem[Sun and Robicheaux(2008)]{Sun08}
B.~Sun and F.~Robicheaux, \emph{New J. Phys.}, 2008, \textbf{10}, 045032\relax
\mciteBstWouldAddEndPuncttrue
\mciteSetBstMidEndSepPunct{\mcitedefaultmidpunct}
{\mcitedefaultendpunct}{\mcitedefaultseppunct}\relax
\EndOfBibitem
\bibitem[Orlando and Reffert(2007)]{Orlando07}
D.~Orlando and S.~Reffert, \emph{arXiv:0709.1546}, 2007\relax
\mciteBstWouldAddEndPuncttrue
\mciteSetBstMidEndSepPunct{\mcitedefaultmidpunct}
{\mcitedefaultendpunct}{\mcitedefaultseppunct}\relax
\EndOfBibitem
\bibitem[D'Ariano \emph{et~al.}(2001)D'Ariano, Lo~Presti, and Paris]{DAriano01}
G.~M. D'Ariano, P.~Lo~Presti and M.~G.~A. Paris, \emph{Phys. Rev. Lett.}, 2001,
  \textbf{87}, 270404\relax
\mciteBstWouldAddEndPuncttrue
\mciteSetBstMidEndSepPunct{\mcitedefaultmidpunct}
{\mcitedefaultendpunct}{\mcitedefaultseppunct}\relax
\EndOfBibitem
\bibitem[Briegel \emph{et~al.}(2009)Briegel, Browne, D\"ur, Raussendorf, and
  Van~den Nest]{Briegel09}
H.~J. Briegel, D.~E. Browne, W.~D\"ur, R.~Raussendorf and M.~Van~den Nest,
  \emph{Nature Phys.}, 2009, \textbf{5}, 19--26\relax
\mciteBstWouldAddEndPuncttrue
\mciteSetBstMidEndSepPunct{\mcitedefaultmidpunct}
{\mcitedefaultendpunct}{\mcitedefaultseppunct}\relax
\EndOfBibitem
\bibitem[De~Pasquale and Facchi(2009)]{DePasquale09}
A.~De~Pasquale and P.~Facchi, \emph{Phys. Rev. A}, 2009, \textbf{80},
  032102\relax
\mciteBstWouldAddEndPuncttrue
\mciteSetBstMidEndSepPunct{\mcitedefaultmidpunct}
{\mcitedefaultendpunct}{\mcitedefaultseppunct}\relax
\EndOfBibitem
\bibitem[Tee(2005)]{Tee05}
G.~J. Tee, \emph{Res. Lett. Inf. Math. Sci.}, 2005, \textbf{8}, 123\relax
\mciteBstWouldAddEndPuncttrue
\mciteSetBstMidEndSepPunct{\mcitedefaultmidpunct}
{\mcitedefaultendpunct}{\mcitedefaultseppunct}\relax
\EndOfBibitem
\bibitem[Lehmberg(1970)]{Lehmberg70}
R.~H. Lehmberg, \emph{Phys. Rev. A}, 1970, \textbf{2}, 883\relax
\mciteBstWouldAddEndPuncttrue
\mciteSetBstMidEndSepPunct{\mcitedefaultmidpunct}
{\mcitedefaultendpunct}{\mcitedefaultseppunct}\relax
\EndOfBibitem
\end{mcitethebibliography}
}

\end{document}